\documentclass[journal,final]{IEEEtranTCOM}

\usepackage{mdframed}
 \definecolor{electricyellow}{rgb}{1.0, 1.0, 0.0}
\usepackage[hidelinks]{hyperref}
\usepackage{cite}
\usepackage{amsmath,amssymb,amsfonts}
\usepackage{algorithmic}
\usepackage{graphicx}
\usepackage{textcomp}
\usepackage{pbox}
\usepackage{multirow} 
\usepackage{booktabs}
\usepackage[normalem]{ulem}
\let\labelindent\relax
\usepackage{enumitem}
\usepackage[keeplastbox]{flushend} 
\usepackage{makecell} 
\usepackage{rotating} 
\usepackage{verbatim} 
\usepackage{amssymb}
\usepackage{amsfonts} 
\usepackage{graphicx} 
\usepackage{algorithm}
\usepackage{algorithmic}
\usepackage{longtable}
\usepackage{supertabular} 
\usepackage{varioref}
\usepackage{cleveref}
\usepackage{cuted}

\newcommand\hlBlue[1]{\textcolor{blue}{#1}}

\usepackage[update,prepend]{epstopdf}
\usepackage{cite}
\usepackage{color}
\usepackage{flushend} 
\usepackage{soul}
\usepackage[caption=false]{subfig}
\usepackage{etoolbox}

\usepackage[absolute,showboxes]{textpos}

\newcommand{\AddFive}{\vspace{5pt}}

\makeatletter
\patchcmd{\@makecaption}
{\scshape}
{}
{}
{}
\makeatletter
\patchcmd{\@makecaption}
{\\}
{.\ }
{}
{}
\makeatother
\def\BibTeX{{\rm B\kern-.05em{\sc i\kern-.025em b}\kern-.08em
    T\kern-.1667em\lower.7ex\hbox{E}\kern-.125emX}}

\usepackage{graphicx}

\begin{document}
	%
	
	
\title{\Huge Hybrid RF/VLC Systems: A Comprehensive Survey on Network Topologies, Performance Analyses, Applications, and Future Directions}



\author{Hisham~Abuella,~\IEEEmembership{Student~Member,~IEEE,}
		Mohammed~Elamassie,~\IEEEmembership{Senior~Member,~IEEE,}
		Murat~Uysal,~\IEEEmembership{Fellow,~IEEE,}
		Zhengyuan~Xu,~\IEEEmembership{Fellow,~IEEE,}
		Erchin~Serpedin,~\IEEEmembership{Fellow,~IEEE,}
		Khalid~A.~Qaraqe,~\IEEEmembership{Senior~Member,~IEEE,}
		and~Sabit~Ekin,~\IEEEmembership{Senior~Member,~IEEE}
		
		
		
\thanks{H.~Abuella and S.~Ekin are with the School of Electrical and Computer Engineering, Oklahoma State University, Oklahoma, USA.}

\thanks{M.~Elamassie and M.~Uysal are with the Department of Electrical and Electronics Engineering, Ozyegin University, Istanbul, Turkey.}

\thanks{Z.~Xu is with the University of Science and Technology of China, Hefei, China.}

\thanks{E.~Serpedin is with the Department of Electrical and Computer Engineering, Texas A\&M University, College Station, Texas, USA.}

\thanks{Q.~A.~Qaraqe is with the Department of Electrical and Computer Engineering, Texas A\&M University at Qatar, Doha, Qatar.}

\thanks{Corresponding author: Sabit~Ekin (e-mail: sabit.ekin@okstate.edu).}
	}

	\maketitle

\begin{abstract} 
 Wireless communications refer to data transmissions in unguided propagation media through the use of wireless carriers such as radio frequency (RF) and visible light (VL) waves. The rising demand for high data rates, especially, in indoor scenarios, overloads conventional RF technologies. Therefore, technologies such as millimeter waves (mmWave) and cognitive radios have been adopted as possible solutions to overcome the spectrum scarcity and capacity limitations of the conventional RF systems. In parallel, visible light communication (VLC) has been proposed as an alternative solution, where a light source is used for both illumination and data transmission.  In comparison to RF links, VLC links present a very high bandwidth that allows much higher data rates. VLC exhibits also immunity to interference from electromagnetic sources, has unlicensed channels, is a very low power consumption system, and has no health hazard. VLC is appealing for a wide range of applications including reliable communications with low latency such as vehicle safety communication. Despite the major advantages of VLC technology and a variety of its applications, its use has been hampered by its cons such as its dependence on a line of sight connectivity. Recently, hybrid RF/VLC systems were proposed to take advantage of the high capacity of VLC links and better connectivity of RF links. Thus, hybrid RF/VLC systems are envisioned as a key enabler to improve the user rates and mobility on one hand and to optimize the capacity, interference and power consumption of the overall network on the other hand. This paper seeks to provide a detailed survey of hybrid RF/VLC systems. This paper represents an overview of the current developments in the hybrid RF/VLC systems, their benefits and limitations for both newcomers and expert researchers. 
\end{abstract}


	\section{Introduction}
	
Wireless communication systems have undergone many changes and developments since their inception. This coincides with the discovery of electromagnetic waves (EM) and wireless telegraph to the present day as advanced technologies such as smartphones, connected vehicles and the Internet of Things (IoT) became widely available.  All these new technologies rely on wireless communication to adapt to the common demand for high bandwidth and data rates. Over the past decades, mobile communications that started with first-generation (1G) and followed by second, third, fourth and fifth generations (2G, 3G, 4G, and 5G). In addition, the wireless fidelity (Wi-Fi) standards for short-range wireless communication (IEEE 802.11) have evolved rapidly especially in terms of data rate, capacity, and medium access methods.

To cope with over-occupied low-frequency bands and provide higher data rates, the idea of millimeter-wave (mmWave) systems was lately introduced. Unfortunately, using radio frequency (RF) systems in the mmWave band has a lot of challenges in terms of channel modeling and transmission power. In addition, using mmWave band is faced with multiple challenges from propagation characteristics of mmWave band (atmospheric and rain attenuation) to beamforming and alignments issues which are discussed in more details in~\cite{Uwaechia_Comprehensive_mmWave_2020_Journal}.
However, channel modeling and measurements for this band were reported by a lot of studies such as~\cite{Rappaport_2014_mmWave_Book}. Despite these challenges, mmWave systems are seen as one of the strongest candidates for 5G systems and some IEEE standards for these systems were introduced as in~\cite{Nitsche_2014_mmWave_ieee_standard} (IEEE 802.11ad for the 60 GHz band). Alternatives to traditional RF systems in terms of mmWave systems are discussed  in~\cite{Rappaport_2013_mmwave,Rappaport_2014_mmWave_Book,Busari__2017_mmWave_MIMO_Survey}.
	
As an alternative to communication systems that operates in the RF band (3 kHz to 300 GHz), the use of visible light (VL) band (400 THz to 800 THz) for wireless communications has been proposed. The idea of using light to transmit a signal has been proposed by Bell and Tainter in 1880 (photo-phone) as discussed in~\cite{Graham_1993_PHOTOPHONE}. However, the idea of transmitting data using a light source was first introduced using a fluorescent lamp in~\cite{Jackson_1998_fluorescent_LAMP_TxRx}. Later, the idea of using the fast switching light-emitting diodes (LEDs) was discussed for the first time in~\cite{Pang_1999_LED_traffic_light_Communcation}. As early as the 2000s in Japan,  the white-LED was used for both illumination and communication by researchers at Keio University \cite{Tanaka_2000_White_LED_Communcation_Japan_Keio}. After this achievement, numerous studies have been published on how to make use of white-LEDs in communication systems. Based on visible light communication (VLC) technology, light fidelity (Li-Fi) was then proposed in~\cite{Haas_2013_LiFi,Haas_2016_WhatisLiFi} to form a small-cell wireless access network where multiple light sources in an indoor environment are used as access points (APs). Furthermore, in 2011, the first IEEE standard related to VLC (802.15.7-2011) was published by IEEE 802.15 working group for Wireless Personal Area Networks. The details of this standard pertaining to data rates, modulation schemes, and dimming mechanisms are discussed in~\cite{Rajagopal_2012_VLC_IEEE_standard}. As a broadens of IEEE standard of STD 802.15.7-2011 to include much more Optical wireless communications (OWC) technologies, STD 802.15.7-2018 has been proposed with six PHY layers categories \cite{IEEE_STANDARD_VLC_2018}. Particularly, this version describes the use of OWC for optical wireless personal area networks (OWPANs) and well-known as the first IEEE standard for Optical Camera Communication (OCC). Actually, OCC  technology represents a promising approach to take advantage of the benefits of VLC in beyond-5G applications and is one of the key technologies for the IoT. It can also be considered as the best compatible with the available infrastructures and can be used in several scenarios/applications. For example, vehicles can exchange safety-related traffic information such as braking performance, accident notifications, speed's and direction's related information. In OCC, with typically low-frame-rate camera detectors, special modulation schemes are needed for non-flickering illumination \cite{Comm_Mag_2018}. OCC is also a mature technology that is clearly described within the IEEE 802.15.7-2018 standard\cite{IEEE_STANDARD_VLC_2018}. Particularly, there are three physical layer (PHY) categories for OCC within the IEEE 802.15.7-2018 standard:

	\begin{itemize}
		\item \textbf{PHY IV:} For outdoor applications with mobility support. Therefore, it can be used for vehicular communication.
		\item \textbf{PHY V:} For commercial cameras with rolling shutter type such as the used cameras in smartphone. Therefore, it can be suitable for indoor applications with small distances.
		\item \textbf{PHY VI:} For screen signage services. Therefore, it can be used for applications such as TV, tablet, smartphone screens.
		\end{itemize}

Since OCC is one of OWC systems that use cameras to receive data rather than photodetectors \cite{Ref_a} and, recently, the use of advanced cameras in smart devices such as mobile phones has been increasing dramatically, the paths for OCC to address several challenges in different applications are opened \cite{Ref_b}. The most common applications are vehicle-to-everything (V2X) communications \cite{Ref_c, Ref_d}, positioning of smartphones and mobile robots (MRs) \cite{Ref_e, Ref_f}, localized advertising, digital signage, and augmented reality (AR) \cite{Survey2_Chowdhury_Optical_2020}. While OCC can be considered as one of the good solutions for long-distance LOS communication links due to several characteristics such as low interference, high SNR, high security, and high stability with respect to non-fixed communication link distances, the data rate is very limited due to the fact that the sampling rate of commercially widely available cameras is not high. Particularly, it ranges from several bits per seconds (bps) to several kilobits per seconds (kbps) that can be achieved by using cameras of 30 frame per second (FPS) \cite{Ref_h, Ref_I, Ref_j}. It is, however, possible to overcome this problem by using high-speed cameras of 1000 FPS \cite{Ref_k}.

While OCC received good attention, VLC has received much more attention. Particularly,  a lot of studies and surveys which discussed the possible applications, advantages, and limitations of these systems were published, see e.g., \cite{Pathak_2015_VLC_Survey,Karunatilaka_2015_VLC_Survey}. In\cite{Wu_2014_VLC_5G_Survey}, the authors presented the limitations and advantages of using VLC as a candidate for 5G systems. In \cite{Pathak_2015_VLC_Survey}, Pathak \textit{et al.} discussed the details of VLC based systems and some of the possible applications like VLC sensing and indoor localization and even using of VLC in vehicle-to-vehicle (V2V) communication. In \cite{Uysal_2015_V2V_VLC_Channel}, Uysal \textit{et al.} discussed the usage of VLC in V2V and the achieved data rate limitations when utilizing LED headlamp used by automotive industry in VLC. In~\cite{Survey4_Chowdhury_Comparative_Optical_2018}, Chowdhury \textit{et al.} provided an overview on optical wireless communication systems and their network architectures and applications.
		\begin{figure}[t!]
				\centering
		\includegraphics[scale = 1]{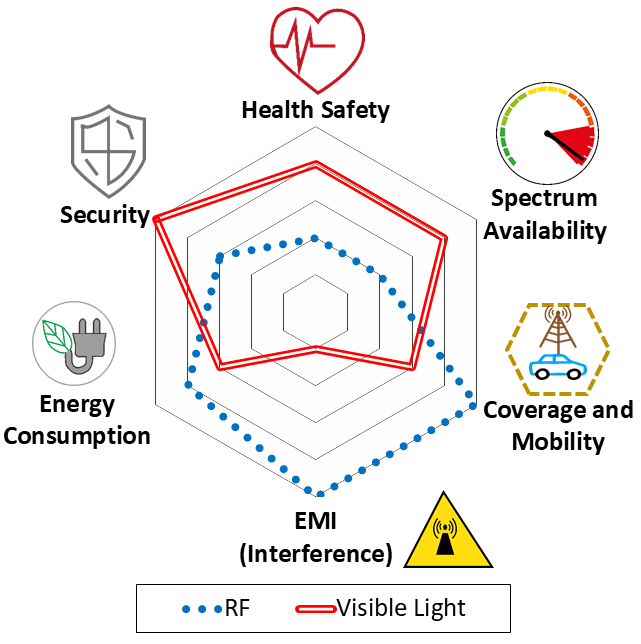}
		\caption{Comparing the pros and cons of VLC \& RF standalone systems.}
		\label{fig:Strength_Weakness_VLC_RF}
		\linespread{-}
	\end{figure}

As for the comparison of the RF and VLC systems, Fig.~\ref{fig:Strength_Weakness_VLC_RF} summarizes the weaknesses and strengths of each of these systems as  standalone networks.

	Based on these, there are several studies investigating how RF and VLC systems can be used as a complementary technology to each other. Analogously, numerous studies have proposed hybrid RF/VLC systems  to overcome the limitations of VLC  by using  RF technology in conjunction with the VLC systems\footnote{The research carried out on hybrid RF/VLC are chronologically ordered and summarized in Table~\ref{table:Hybrid_Ref_years}.}.

	To our best knowledge, the only tutorial discussing hybrid RF/VLC systems was published in~\cite{Ayyash_2016_Coexistence_Hybrid_Magazine,Chowdhury_2014_Cooperative_Journal}. Ayyash \textit{et al.} in~\cite{Ayyash_2016_Coexistence_Hybrid_Magazine} discuss the coexistence opportunities of Wi-Fi and Li-Fi systems, the opportunities that these systems present for off-loading of Wi-Fi systems and the challenges faced by these wireless heterogeneous networks (HetNets) as a future solution in 5G systems. Chowdhury \textit{et al.} in~\cite{Chowdhury_2014_Cooperative_Journal} discuss the possibility of using a hybrid  RF and optical wireless network to meet the 5G requirements. They provide a brief overview of optical wireless networks and the possible hybrid network architectures and describe the possible research directions and improvements for hybrid RF/VLC networks. Moreover, multiple surveys discussing the improvements of VLC and Li-Fi networks have discussed the idea of hybrid RF/VLC networks as in~\cite{Survey1_Obeed_Optimizing_2019,Survey2_Zhuang_Positioning_2018,Survey3_Li_Optimization_2018,Feng_Applying_2016_Survey}. 
	In~\cite{Survey1_Obeed_Optimizing_2019}, Obeed \textit{et al.} focus on the VLC networks and how to optimize them for downlink applications. The authors present the challenges facing VLC networks and proposed solutions by literature to maximize the benefit from VLC networks in terms of sum rate, fairness, energy efficiency, secrecy rate, and harvested energy. However, integration of VLC with current existing RF systems and how to utilize them together are not covered.
    In~\cite{Survey3_Li_Optimization_2018}, Li \textit{et al.} focus on the design of VLC networks system level from network centric to user centric (UC) design principle.  Moreover, the survey focuses on a radically new UC design philosophy and discuss the visible light communication link structure and the unique characteristics. However, this study does not explain how RF and VLC can be used together in different network designs. It also does not discuss the performance analysis of different hybrid systems, their applications and environments.
    In~\cite{Feng_Applying_2016_Survey}, Feng \textit{et al.} presented a magazine article which discusses the VLC networks and their main architecture. Moreover, the features of VLC networks  make them a good candidate to play an important role in 5G networks. However, this article lacks important aspects of how hybrid systems are designed and used in different applications.
    
	A survey was published on arXiv~\cite{Survey1_Wu_Hybrid_2020} which focuses more on optimizing the Li-Fi/ Wi-Fi network parameters (user behavior modeling, interference management, handover, and load balancing) but did not focus on the general RF/VLC systems. 
	
	In addition, a general overview of the optical wireless hybrid systems and current research trends and issues in network layer level is presented in \cite{Survey2_Chowdhury_Optical_2020}. In \cite{Survey2_Chowdhury_Optical_2020}, authors focus on optical wireless hybrid systems, which include VLC and OWC. The authors have introduced the difference between different light-based technologies such as VLC, Li-Fi, OCC and free space optical (FSO) communications. Moreover, the authors presented different applications that cover hybrid systems. The author presented system models to show the different application where hybrid systems can be used.
    After the publication of~\cite{Chowdhury_2014_Cooperative_Journal},  a large number of studies have been reported on the design and performance analyses of these systems as described in chronological order in Table~\ref{table:Hybrid_Ref_years}.
       
    In addition some studies discussed the use of hybrid RF/VLC in high level applications. Liu \textit{et al.} in~\cite{Liu_Hybrid_2012_Conf} presented an intelligent transportation hybrid system that utilizes the directionality of visible light to send a lane specific code for vehicles to decode a message sent by RF link which improves the effective communication area.
    Yang \textit{et al.} in~\cite{Yang_Integration_2020_Journal} discussed the idea of integrating visible light communication and positioning to assist in 5G networks for IoT devices where macrocell and picocell provide coverage and reliability in RF spectrum, and optical attocell provide the high-speed transmission and high-accuracy positioning services.
    
	\textbf{This survey focuses on  achieving  the following goals:} 
	\begin{enumerate}
	
    \item Highlighting the design aspects of VLC systems such as channel modeling, system performance, advantages, and limitations. 

    \item Highlighting the RF technologies that are used in hybrid systems and their corresponding channel models. 
    
    \item Offering readers a general understanding of hybrid RF/VLC systems, their history, and the desired goals intended to accomplish. 

    \item Pointing out the contributions of the major hybrid RF/VLC studies according to their network topology and performance analysis.

    \item Outlining the challenges in the development of hybrid RF/VLC systems, future research directions, and possible promising applications.
    
    \item Comparing all of the existing studies on hybrid RF/VLC systems in terms of the employed technologies, network topologies  and main features.
    
    \item Pinpointing research challenges, future research directions, and possible applications.
		
	\end{enumerate}

The rest of the paper is organized as follows. VLC and RF systems are presented in Section~\ref{sec:VLC_RF}. Hybrid RF/VLC systems, network topologies, performance analysis studies, system-level simulations, and implementations are discussed in Section~\ref{sec:Hybrid_sys}. Research directions and emerging applications of hybrid RF/VLC systems are presented in Section~\ref{sec:Research_Directions}. This survey ends with concluding remarks on the potential of hybrid RF/VLC systems. Nomenclature and key symbols used in the paper are summarized at the end of the paper.

	\section{VLC and RF Technologies}
	\label{sec:VLC_RF}
	
	In this section, we provide an overview of the VLC and RF systems highlighting the used modulation techniques and channel properties. We further compare these two systems motivating the need for developing hybrid RF/VLC systems.
	
	\phantomsection
	\subsection{VLC Systems}
	
    The main requirement for a VL source to be used in a communication link is that the light intensity should be modulated at a rate higher than 200 Hz~\cite{Rajagopal_2012_VLC_IEEE_standard} to avoid any flickering effects to the human eyes. Despite this minor requirement,  VLC research has shown that high data rates can be achieved (nearly 100 Mbps in IEEE 802.15.7~\cite{Rajagopal_2012_VLC_IEEE_standard}).  
	
	In addition to academic/theoretical VLC research activities, industry interest in VLC has sparked related standardization activities in this emerging market. For example, the VLC consortium (VLCC) initiated the standardization activities and proposed two standards which are accepted by Japan Electronics and Information Technology Industries Association (JEITA). The two proposed standards by VLCC are VLC system standard (CP-1221) and visible light ID standard (CP-1222). The latter has been updated to visible light beacon system standard (CP-1223). CP-1223 describes the unidirectional VLC system for multimedia applications with supported wavelengths in the range of 380 - 780 nm and data rate around 4.8 kbps by the use of inverted 4 pulse position modulation (I-4PPM). Similarly, IEEE introduced IEEE Standard 802.15.7, which is defined for short-range communication. The sufficient data rates for supporting audio and video multimedia services are specified. This standard defines two layers which are physical layer (PHY) and medium access control layer (MAC). They have defined three types of PHY layers based on the supported frequency band, optical clock rate and data rate. The first PHY (PHY I) is defined for outdoor with low data rate applications. Particularly, it can support up to 266.6 kbps. The second PHY (PHY II) is defined for indoor with moderate data rate applications. Modulation schemes of on-off keying (OOK) and variable pulse position modulation (VPPM) up to 96 Mbps are supported. The third PHY (PHY III) is defined for application with multiple light sources/detector. Modulation scheme of color-shift keying (CSK) up to 96 Mbps is supported. On the other hand, the MAC layer handles accessing to PHY layer. The dedicate portions, also called guaranteed time slots (GTSs), are assigned by a coordinator device during contention free period
\cite{VLC_Book_2017, CP_Standard_Conf, IEEE_STANDARD_VLC_2018,Yuwei_Standard}.

    Despite the fact that the optical channel of visible light is in the order of THz, the bandwidth of commercial LEDs is limited. It has been shown that the LED’s modulation bandwidth is in the order of MHz to hundreds of MHz \cite{Wang_VLC_BOOK}. This makes it a bottleneck for achieving high-rate transmission. Different efforts have been conducted to address this. The first optical transmitter that was adopted is phosphor-converted LED. The bandwidth of a phosphor-converted LED is limited because of its slow response. A post-equalization circuit which consists of one active equalizer and two passive equalizers was then proposed to extend the bandwidth up to 150 MHz \cite{Li_Photon_Lett}. Notice the fact that if higher bandwidth LEDs are employed, the bandwidth can be significantly increased. For example, 1 Gbps was reported by using micro LEDs as transmitters \cite{McKendry_Photon_Let}.
    
    Different experiments have been conducted in the literature comparing line-of-sight(LOS) with non-LOS (NLOS) Links. In \cite{Eldeeb_Conf}, they have carried out experimental studies where the measurements have been done in an empty room for different scenarios including LOS and NLOS cases. They have considered frequency sweeping technique in their measurements by using of vector network analyzer (VNA) in an effort to obtain channel impulse response (CIR). While the reflected rays can still be detected in NLOS scenarios, the channel gain severely dropped (see  \cite[Fig. 6]{Eldeeb_Conf}). Similar observations can be seen in \cite{UYSAL_IEEE_STANDARD} where a commercial optical and illumination design software is used for obtaining CIR. Moreover, uplink in VLC systems can be considered as one of the major challenges. Furthermore, the advancements in the LED industry are still hard to support the user equipment (UE)'s side with a VLC transmitter. Moreover, VLC is sensitive to interference caused by other light sources which can be clearer in daylight or outdoor scenarios. Also, the dimming mode for using VLC in low light scenarios can be an issue although an advanced modulation technique (enhanced unipolar orthogonal frequency division multiple access (eU-OFDM)~\cite{Tsonev_2015_DimmingMode_VLC_Sol}) has been proposed for Li-Fi systems to operate in the very low light intensity scenarios while keeping a reasonable data rate. An important design consideration for VLC networks is the LED connectivity to the internet source.  Typically,  VLC networks present a large number of APs of LED arrays which make internet connectivity challenging. 
    Most of these limitations have been recognized in literature and solutions have been proposed to resolve them \cite{Haas_2018_5G_LiFi}. We will outline the main solutions later in this survey.

\begin{figure}[!t]
		\includegraphics[clip,trim=6 18 13 -10, scale=0.45]{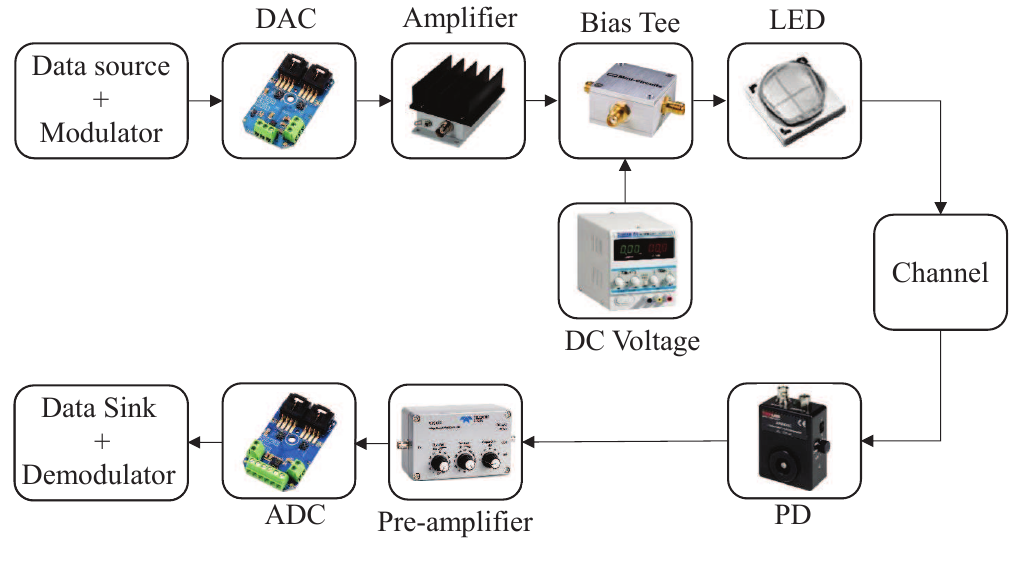}
		\centering
		\caption{Block diagram of the basic structure of the VLC system.}

		\label{fig:VLC_system_Model}
	\end{figure}

    Next, we will focus on introducing the basic VLC system components, modulation techniques and channel properties that help in developing a robust hybrid RF/VLC system with respect to the limitations outlined above.
     Fig.~\ref{fig:VLC_system_Model} depicts the basic structure of the VLC system. We will later outline the main differences between the RF systems and VLC systems. It is clear that the base-band algorithms will not change a lot except for the fact that LEDs and photo-detectors (PDs) operate in the non-negative voltage range, unlike the RF antennas. At the transmitter side, an LED is used instead of an antenna in the RF system. This LED needs two inputs: an analog signal and a direct current (DC) supply to drive the LED. Additionally, like in the RF-based systems, a digital to analog converter (DAC) is used to convert the digital signal into an analog signal after getting the modulated data from the base-band modulator. At the receiver side, a PD is used instead of a receiving antenna and the signal is amplified by using an electric amplifier, sometimes adding some optical lens and filters to improve the field of view (FOV) and gain of the receiver. After amplifying the signal, an analog to digital converter (ADC) is used to convert the analog signal into a digital signal that is to be processed by the base-band digital demodulator.

		\begin{table*}
		\linespread{1.2}
		\footnotesize
		\begin{center}
			\caption{Comparison of different VLC receivers in terms of cost, data rate, sensitivity, size, power consumption and availability in current UE.}
			\label{table:Comparison_VLC_RX}
			\begin{tabular}{|>{\centering\arraybackslash}m{3.2cm}|>{\centering\arraybackslash}p{2.6cm}|>{\centering\arraybackslash}m{2.6cm}|>{\centering\arraybackslash}m{4.6cm}|}
				\cline{2-4}
				\multicolumn{1}{ c |}{} & \textbf{Photo-detector} & \textbf{Solar-cell} & \textbf{Imaging-sensor}\\
				\hline
				\raggedright{Cost}  &  Moderate & Low & High \\ \hline
				\raggedright{Data rate}  & High & Moderate & Moderate \\  \hline 
				\raggedright{Sensitivity} & Sensitive & High Sensitivity  & Low Sensitivity\\  \hline
				\raggedright{Size} & Small & Large & Moderate \\ \hline 
				\raggedright{Power consumption} & Moderate & Low & High\\ \hline 
				\raggedright{Availability in current UE} & No & Rare & Available with modification needed\\ \hline
			\end{tabular}
		
		\end{center}
	\end{table*}
	
	\phantomsection\subsubsection{VLC Front-Ends (Transmitter/Receiver)}
There are two major types of structures for colored/white LEDs used in general lighting. The first type consists of a blue-colored LED with a phosphor layer coated on top of it. When an electric current is applied to the LED, light is emitted from that LED and part of it is absorbed by the phosphor to generate the second color. The combination of these two colored lights results in colored/white light. Another type of LED is produced by mixing lights from three primary colored chips (RGB). Three chips emit each color simultaneously and at the output, the required colored/white light is produced. The phosphor white LED has the advantage of low cost. However, the nature of phosphor light conversion makes it unsuitable for high-speed data communication due to phosphorous response time. In other words, white phosphor LEDs (WPLEDs) exhibit a limited bandwidth of few MHz while RGB LEDs present a  higher bandwidth with an order of magnitude higher than WPLEDs \cite{ VLC_Book_2017, VLC_Book_2019}. Furthermore, RGB-LEDs are suitable for color shift keying (CSK) modulation technique. This will be discussed in more detail in the VLC modulation methods subsection. 
	
		\begin{figure}[t!]

		\includegraphics[clip,trim=25 23 71 5, scale=.5]{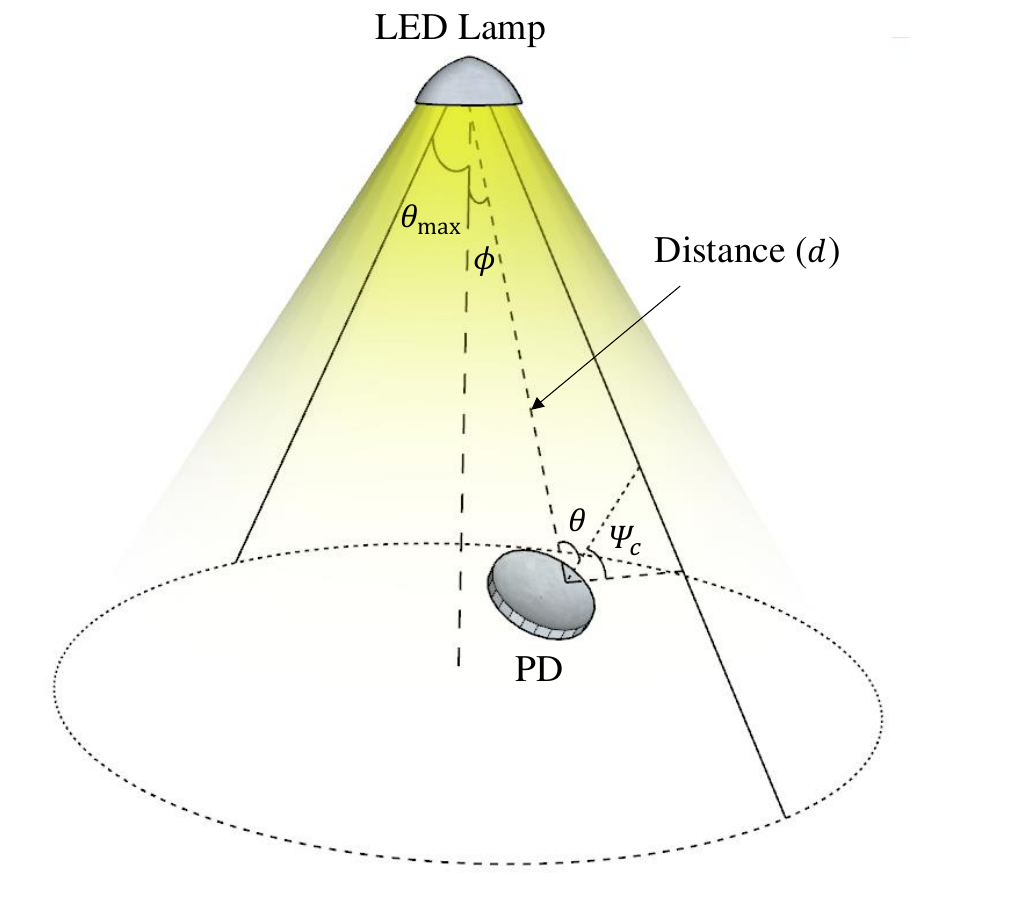}
		\centering
	
		\caption{LED based LOS channel model \cite{Cui_2010_LOS_VLC_Demo}, where a VLC transceiver system model illustrating the parameters of VLC channel models is presented.}
		\label{fig:VLC_Link_Level_system}
	\end{figure}

	On the receiver side, there are three types of receivers that can be used:
	\begin{itemize}
		\item \textbf{ Photo-detector (photo-diode) (PD))}:
		
		It is a semiconductor device that converts the light energy to a voltage difference and it can support very high data rates.
		
		\item \textbf{ Solar-cell\footnote{Solar-cells, typically, detect light and other electromagnetic radiation near the visible range such as infrared.}}:
		
		It is a larger version of a photo-detector used to collect as much light energy as possible.
		
		\item \textbf{ Imaging-sensor}:
		
		It is an array of photo-detectors that can be used to improve the data rates.
	\end{itemize}
	
	Table~\ref{table:Comparison_VLC_RX} illustrates  a comparison between the three methods for receiving the signal in VL-based systems in terms of cost, data rate, sensitivity, size, power consumption and availability in current UE.
		
In this survey, it is assumed that the VLC systems use normal blue LEDs with a phosphor transmitter and photo-detector receivers unless stated otherwise.
    
    \subsubsection{VLC Path Loss Models (Channel Models)}
        Several works have been done on channel modeling and characterization for indoor optical link \cite{ Gfeller_and_Bapst_Channel_1979_Journal, Barry_1993_channel_model_OWC, Perez_Channel_1997_journal, Lopez_Channel_1998_journal, Carruthers_Channel_1997_journal, Carruthers_Channel_2002_journal, Hayasaka_Channel_2007_journal, Perez_channel_2002_journal} and outdoor \cite{Karbalayghareh_VVLC,Al_Kinani_VVLC2} optical link.
 For example, assuming single reflections, Gfeller and Bapst in \cite{ Gfeller_and_Bapst_Channel_1979_Journal } have proposed the first propagation model. Assuming multiple reflections, Barry \textit{et. al.} in \cite{Barry_1993_channel_model_OWC} have proposed a recursive simulation model. While the previous models were proposed for IR, in 2011,  Lee  \textit{et. al.} have extended the channel models of IR and produced the first indoor channel model for VLC systems \cite{ Lee_Channel_2011_journal}.
    
        For indoor VLC systems as shown in  Fig.~\ref{fig:VLC_Link_Level_system}, the transmitter, which is based on LEDs, is generally modeled by a Lambertian pattern\cite{Gfeller_and_Bapst_Channel_1979_Journal,Barry_1993_channel_model_OWC,Kahn_article_1997}. The channel gain based on LED with Lambertian pattern that considers LED beam solid angle, LED beam maximum half-angle, the angle between the source-receiver line and beam axis, and the angle between the source-receiver line and receiver normal is adopted by most papers that consider hybrid RF/VLC. VLC path loss models can be summarized as follow:
	
	\begin{itemize}
		\item \textbf{Lambertian model:}
		
		For indoor VLC, the channel gain of LOS is generally modeled by the Lambertian emission and given by \cite{Kahn_article_1997}
		\begin{equation}\label{eq:Lambertian_Channel1}
		\begin{split}
		H_v = &\frac{(m+1)A_R}{2\pi d^2} T_s(\theta) g(\theta) \cos^m(\phi)\\ &\times \cos(\theta),\qquad \forall\theta < \phi_{\text{max}},
				\end{split}
		\end{equation}
		
		where $A_R$ is the optical detector size, and $d$ is the distance between the receiver and the transmitter.  $ \phi $ and $ \theta $ stand for irradiance and incidence angles, respectively.
		In addition,  $\phi_{\text{max}}$ represents the semi-angle at half-power of the LED, $m$ is the order of the Lambertian model and is given by $m=-\frac{\ln(2)}{\ln(\cos\phi_{\text{max}})}$, $T_s(\theta)$ is the gain of the optical filter, and $g(\theta)$ is the concentrator gain, which is assumed to be a constant depending on the concentrator design.  
		
		\item \textbf{Simple path loss model:}
		
		In the simple model (similar to the RF path loss model \cite{Rappaport_2014_mmWave_Book}), the power-distance relation is modeled as follows
		\begin{equation}
		\label{eq:Simple_Channel}
		H_v=K\left[d\right]^{-\varsigma },\qquad    \forall~d > 1,
		\end{equation}
		
		where $H_v$ is the optical channel gain, $K$ is a constant that represents all the gains and the transmitted power and $\varsigma$ denotes  the channel path loss exponent which usually depends on the channel environment.
		This simple model assumes that the channel only has a path loss factor and a gain which is dependent on the system gains and  interference from the environment. It is used by the studies where the VLC system is assumed  static and only the distance of the LOS is changing without a change in the irradiance and incidence angles as in ~\cite{Ray_tracing_Eldeeb_Uysal_Conf_2019,Abuella_Vildar_2019_Journal,Channel_VLC_Ref2}.
			\end{itemize}
		  
	For outdoor, different VLC channel models are available in the open literature either via simulation software programs \cite{Elamassie_Fog,Karbalayghareh_VVLC}, analytical means \cite{Al_Kinani_VVLC,Qingshan_VVLC,Al_Kinani_VVLC2} or experimental methods \cite{Tseng_VVLC,Pengfei_VVLC,Bassam_IWOW}. In those models, either high beam headlamps \cite{Elamassie_Fog,Karbalayghareh_VVLC} or low-beam headlamps \cite{Al_Kinani_VVLC,Qingshan_VVLC,Al_Kinani_VVLC2,Tseng_VVLC,Pengfei_VVLC,Bassam_IWOW} were considered as transmitting light sources. For instance, with a high beam headlight, a V2V channel model for perfect alignment case was proposed in \cite{Elamassie_Fog} for distances up to 20 m using non-sequential ray tracing. This model was extended in \cite{Karbalayghareh_VVLC} considering longer link distances and lateral displacements. While high beam has been widely considered since it allows further illumination and hence communication distances, there are several regulations\cite{Driving_Reg1} preventing the use of high beam headlights in several scenarios \cite{Driving_Reg2}. These limitations motivate researchers, recently, to use low-beam headlights for vehicular VLC connectivity \cite{Bassam_IWOW}. While \cite{Al_Kinani_VVLC,Qingshan_VVLC,Al_Kinani_VVLC2} considered Lambertian channel model, a model based on measurements of low-beam tungsten-halogen bulbs was introduced in \cite{Pengfei_VVLC}. Since patterns of the LED-based headlight are different from those of the traditional halogen headlight, the first experimental model for LED-Based headlight was proposed in \cite{Bassam_IWOW}.

	\subsubsection{VLC Modulation Methods (PHY)}
	The typical candidates for the front-end devices in VLC systems are incoherent LEDs and laser diodes (LD) because of their low costs. Due to their physical characteristics, information can be transmitted by modulating the intensity of the light. Therefore, the transmitted signal should be a unipolar non-negative real-valued signal. Such a VLC technique of modulation/demodulation is referred to as  IM/DD (Intensity Modulation/ Direct Detection).
	For IM/DD, some techniques can be applied in, relatively, straightforward manner, e.g., pulse width modulation (PWM), pulse position modulation (PPM), on-off keying (OOK), and pulse amplitude modulation (PAM). As the data rate increases, these modulations begin to suffer from the effects of intersymbol interference (ISI) as a result of frequency selectivity. Therefore, a more suitable modulation technique is needed. A typical candidate is orthogonal frequency division multiplexing (OFDM). Not only OFDM has a flat fading channel for each subcarrier but also allows for an adaptive bit and power loading which leads to optimal utilization of available resources. Furthermore, OFDM includes simple equalization in the frequency domain with single-tap equalizers in addition to its ability to avoid frequency distortion due to flickering. It should be, however, noted that the typical OFDM signals are bipolar complex-valued. Therefore, the standard OFDM must be modified in order to make it suitable for IM/DD.

	On the other hand, the use of light, simultaneously, for illumination and data communication purposes poses some challenges that require consideration in implementing the VLC system. The two main challenges are dimming support and flicker mitigation. Typically, the lighting fixture is equipped with dimming control that allows users to control the level of brightness they prefer, while flicker is observed by the human eye as a result of continuous switching between on and off during data transmission.

The main modulation schemes presented in the literature can be itemized as follows:
\begin{itemize}
		\item  \textit{On-off keying}: OOK is the simplest form of modulation that represents data in VLC as, typically, the presence or absence of light. In this form, the presence of light represents a binary one, while the absence represents a binary zero. Some more sophisticated OOK schemes vary the duration of presence and absence~\cite{Grubor_2007_OOK,Minh_2008_OOK,Vuvcic_2010_OOK,Fujimoto_2013_OOK}.
		
		\item  \textit{Pulse modulation}: Pulse modulation is a simple modulation scheme where the transmitted signal is presented in form of pulses~\cite{Ntogari_2011_PulseMod,Bai_2010_PulseMod,Noshad_2014_PulseMod,Noshad_2012_PulseMod}. It is classified into three major types:
		\begin{itemize}
		\item  Pulse-width modulation/Pulse-duration modulation
		\item  Pulse-amplitude modulation
		\item  Pulse-position modulation
        \end{itemize}
		
		\item  \textit{Orthogonal frequency division multiplexing}: OFDM is a multi-carrier technique, where the available frequency band is divided into many small bands by use of orthogonal sub-carriers \cite{Mesleh_2011_OFDM_perf_OWC_HaraldHaas,Armstrong_2009_OFDM,Afgani_2006_VLC_OFDM,Armstrong_2006_VLC_OFDM,Elgala_2007_VLC_OFDM,Tsonev_2014_VLC_OFDM_demo}.
		
		\item  \textit{Color shift keying}: CSK is a VLC modulation scheme, where the data is transmitted by the use of different colors with the same intensity \cite{Drost_2010_color_shift_keying,Monteiro_2014_color_shift_keying,Monteiro_2012_color_shift_keying}.
		
\end{itemize}

	\subsubsection{VLC Multiple Access Schemes}
	In VLC, different multiple access methods that allow more than two users/nodes to be connected to the same transmission medium and transmitted over it have been proposed such as:

\begin{itemize}
		\item  \textit{Carrier sense multiple access (CSMA)}: a user/node tries firstly to determine if another transmission is in progress before setting up its transmission by the use of carrier-sense mechanism (CSM). If CSM sensed a current transmission, user/node waits for the current transmission to be ended~\cite{Mai_2017_CSMA_VLC_WiFi}.
		
		\item  \textit{Orthogonal frequency division multiple access (OFDMA)}: OFDMA is a modified version of the OFDM where multiple access is achieved by assigning different subcarriers to different users/nodes. This allows simultaneous transmission from several users/nodes~\cite{Ling_2018_OFDMA_VLC}.
		
		\item  \textit{Code-division multiple access (CDMA)}: CDMA is a channel access scheme where several users/nodes can send data simultaneously over a single VLC channel. This allows several users/nodes to share the same band of frequencies without too much interference between them by the use of a special coding scheme where each user/node is assigned a special code~\cite{Qiu_2017_CDMA_VLC}.

        \item  \textit{Non-orthogonal multiple access (NOMA)}: in NOMA, users/nodes share the available time and frequency resources simultaneously, which leads to low latency and better spectral efficiency \cite{Ref_RF_NOMA,NOMA_ref_reviewer_Hanzo_2020}. NOMA systems can be accomplished in two domains, i.e., power or code domains. The most commonly used one is NOMA with a power domain where different power levels are assigned to different users/nodes. On the other hand, in the NOMA scheme with code-domain, multiplexing is accomplished by the use of spreading sequences similar to CDMA technology. 
        
        The main differences between NOMA-VLC and NOMA-RF are the achievable capacity and the realization of successive interference cancellation (SIC). Since the classical Shannon's capacity does not work for optical systems, the exact capacity of a VLC system is still unknown. Alternatively, upper and lower bounds on optical capacity are derived in the literature \cite{Capacity_Optical_a,Capacity_Optical_b,Capacity_Optical_c,Capacity_Optical_d}. It has been shown that the gap between the lower bound presented in \cite{Capacity_Optical_d} and the exact capacity can be neglected for high SNR. This simply requires scaling the SNR in Shanon's capacity by a constant of $\exp\left(1\right)/2 \pi$. On the other hand, CSI is required at both the receiver and the transmitter sides for splitting the transmit power with suitable coefficients among users that allows the practical realization of SIC. This requires updating CSI at both transmitters and receiving nodes with a rate larger than the frequency of channel changes which is in RF more than VLC due to the fact that the VLC channel remains unchanged most of the time \cite{Marshoud_NOMA_VLC}. 
        
\end{itemize}
	
	It should be further noted that the access method can be a part of multiple access protocol (MAP) and control mechanism (CM), which is known as medium access control (MAC). MAC deals with issues such as assigning channels to different users/nodes.

	\subsection{RF Systems}
    Since RF systems are old and well-established technology, researchers have used multiple standards and bands to improve their performance. Most of the proposed solutions employ Wi-Fi (microWave) technology. Conventional Wi-Fi operates in frequency ranges that are below 6 GHz\footnote{There are several RF ranges for Wi-Fi communications: 900 MHz, 2.4 GHz, 3.6 GHz, 4.9 GHz, 5 GHz, 5.9 GHz, and 60 GHz bands. However,  communication with 60 GHz is limited to a few meters and can not pass through walls compared to conventional Wi-Fi frequencies.}. In this frequency range, the data rate is in the range of several 100 Mbps. For achieving several Gbps, some studies explored the possibility of using mmWave systems instead of old microWave systems. The different RF systems used in RF/VLC systems studies can be classified in the following categories:

	    \subsubsection{Wi-Fi}Wi-Fi is a wireless communication technology that refers to the IEEE communications standards for wireless local area networks (i.e., IEEE 802.11) and was created in 1997. This technology which is based on the direct sequence spread spectrum (DSSS) and frequency-hopping spread spectrum (FHSS) uses radio signals that allow accessing the internet while moving from one place to another via high-speed network connections. IEEE 802.11 supports up to 2Mbps data rates. Different versions of Wi-Fi (i.e., different standards) have been proposed: IEEE 802.11a which works on the 5 GHz band and yields a maximum of 54Mbps is based on OFDM and was created in 1997. IEEE 802.11b  works on the 2.4 GHz band and offers a maximum of 11Mbps and was created in 1999.  IEEE 802.11g  works on the 2.4 GHz band,  provides a maximum of 54Mbps and was created in 2003. IEEE 802.11n works on the 2.4 GHz and 5 GHz bands and was created in 2009. This standard supports multi-channel with a maximum of 150Mbps/channel. In addition, IEEE 802.11ac  was invented in 2014, and this standard increases the data rate up to 1300 Mbps  \cite{Gast_2013_WiFi_Gigabit_Book,Wang_2018_mmWave_Surv}.
    
        \subsubsection{mmWave} Most commercial radio communications including Wi-Fi work in a narrow band of the RF spectrum (i.e., 300 MHz-3 GHz).  However, the part of the RF spectrum above 3 GHz is mostly unexploited for commercial applications. Recently, there is a huge interest in utilizing this range. For example, the range  3.1-10.6 GHz has been proposed for high data rate connectivity in personal area networks. The range  57-64 GHz is used to provide data rates at the order of Gbps for short-range local area networks. Furthermore, the range of 28-30 GHz has been proposed for local multipoint distribution services.
        Millimeter waves can support high data rates at the order of  Gbps but are severely affected by the absorption caused by oxygen molecules and water vapors from the atmosphere. Since oxygen absorbs EM waves at around 60 GHz, the frequency range  57-64 GHz can experience huge attenuation on the order of 15 dB/km. On the other hand, the range  164-200 GHz is severely affected by the concentration of water vapors in the atmosphere and may be subject to attenuation on the order of tens of dB/km \cite{Gupta_2015_Survey_5G,Pi_2011_millimeter,Wei_2014_mmWave}.
        
        \subsubsection{Dedicated Short Range Communications (DSRC)} DSRC is a wireless communication standard that is designed specifically for short and  medium communication ranges. It is used mainly for vehicular communications, i.e., V2V, vehicle-to-infrastructure (V2I) and infrastructure-to-vehicle (I2V) communications. DSRC systems exploit microwaves in the ranges of  5.805-5.815 GHz and 5.795-5.805 GHz \cite{ Goel_2008_DSRC_Book }. The first version of it has been proposed in 1999 \cite{Bitam_2017_DSRC_Book}. In 2003, an improved version referred to as the American Society for Testing and Materials (ASTM)-DSRC standard has been proposed \cite{Bitam_2017_DSRC_Book}.
        DSRC is not just for PHY and MAC layers, it is, actually, a complete communication protocol. The list of DSRC IEEE standards is given in \cite{Kurihara_IEEE_STANDARD_P1609,Vivek_Implementation_IEEE_1609}:
            
        \begin{itemize}
            \item {IEEE 802.11p-PHY Layer:}
            \item {IEEE P1609.1:} Standard for Wireless Access in Vehicular Environments (WAVE) - Resource Manager. 
            \item {IEEE P1609.2-Security Layer:} Standard for Security Services for Applications and Management Messages. 
            \item {IEEE P1609.3-Network and Transport Layers:} Standard for Networking Services. 
            \item {IEEE P1609.4-MAC Layer:} Standard for Multi-Channel Operation. 
            \item {IEEE P1609.11:} Standard for WAVE-- Over-the-Air Electronic Payment Data Exchange Protocol for Intelligent Transportation Systems (ITS)
            \item {IEEE P1609.12:} Standard for WAVE - Identifier Allocations
        \end{itemize} 
	
	Table~\ref{table:RF_Channel_Models} shows how different hybrid RF/VLC systems studies employ different RF channel models based on the RF system assumed in each study.
	
	\begin{table}[t!]
		\linespread{1.15}
		\centering
		\footnotesize
		\begin{center}
		\caption{Utilized RF wireless  technologies in hybrid RF/VLC systems.}
			\label{table:RF_Channel_Models}
			\begin{tabular}{|>{\centering\arraybackslash}m{4.5cm}|>{\centering\arraybackslash}p{3.5cm}|}
				\hline
				\textbf{RF Channel Model} &   \textbf{References} \\
				\hline 
			\end{tabular}
			\begin{tabular}{|>{\arraybackslash}m{4.5cm}|>{\arraybackslash}p{3.5cm}|}
				{{Wi-Fi (2.4 GHz/5 GHz)}} & \cite{Ma_Location_Conf,Khreishah_2018_jour_Energy_Efficient,Wu_2018_Jour_adap_resou_opti,Li_2018_Conf_BW_Aggreg,Wentao_2017_Design_Journal,Wang_2018_Learning_Journal,Wu_2017_Access_Journal,Wang_2017_Optimization_Journal,Bao_2017_Visible_HetNet_Journal,Zenaidi_2017_Achievable_Conf,Wu_2017_Access_Conf,Rakia_2016_Optimal_DualHop_Journal,Shao_2016_Delay_Journal,Rakia_2016_DualHop_VLC_RF_Conf,Yan_2016_combination_Conf,Wu_2016_Two_stage_Conf,Duvnjak_2015_Heterogeneous,Shao_2015_Design_Journal,Wang_2015_Efficient_Journal,Li_2015_Cooperative_Journal,Wang_2015_Dynamic_Journal,Chowdhury_2014_Cooperative_Journal,Shao_2014_Indoor_Conf,Rahaim_2011_hybrid_Conf}  \\ \hline
				WINNER II Channel Model (2-6 GHz)  & \cite{Tabassum_2018_Coverage_Journal,Zhang_2018_Energy_Journal,Kashef_2017_Transmit_PLC_Journal,Kafafy_2017_Power_Conf,Marzban_2017_Beamforming_Conf,Kashef_2016_Energy_Journal,Kashef_2016_Impact_PLC_Conf,Kashef_2015_Achievable_Conf,Stefan_2014_Hybrid_Conf}  \\  \hline 
				mmWave (60 GHz)  & \cite{Obeed_2018_Jour_power_allocation,Basnayaka_2015_Hybrid_Conf,Wang_2015_Distributed_Conf,   Wu_2016_Two_stage_Conf,Obeed_2017_Joint_Conf}  \\  \hline 
				DSRC (5.9 GHz)  & \cite{Bazzi_2016_Visible_Journal}  \\  \hline 
			\end{tabular}\AddFive
		\end{center}
	\end{table}

		\begin{table}[t!]
		\linespread{1.15}
	
		\footnotesize 
		\begin{center}
			\caption{Comparison of VLC and RF technologies that explains the main motivating factors behind hybrid RF/VLC networks.}
			\label{table:Comparison_VLC_RF}
			
			\begin{tabular}{|>{\centering\arraybackslash}m{2.9cm}|>{\centering\arraybackslash}p{2.3cm}|>{\centering\arraybackslash}m{2.3cm}|}
				\cline{2-3} 
				\multicolumn{1}{ c |}{}& \textbf{VLC} & \textbf{RF} \\
				\hline 
				\raggedright{Range}  &  Low (Up to 100 m) & High (Up to 1km)\\ \hline
				\raggedright{Environment dependency} & Sensitive & Moderate \\  \hline 
				\raggedright{Ambient light} & Sensitive & Not affected \\  \hline
				\raggedright{EMI} & No & Yes \\ \hline
				
				\raggedright{Band license} & Unlicensed & Licensed/Unlicensed \\ \hline
				\raggedright{Cost} & Low & High \\ \hline
				\raggedright{Size} & Small & Large \\ \hline 
				\raggedright{Power consumption} & Low & High \\ \hline 
			\end{tabular}
		\end{center}
	\end{table}

	\subsection{Comparison of RF and VLC Technologies}
	
In this section, the comparison of RF and VLC technologies (Table~\ref{table:Comparison_VLC_RF} \cite{Uysal_2015_V2V_VLC_Channel,Ayyash_2016_Coexistence_Hybrid_Magazine}) is presented.

    As shown in Table~\ref{table:Comparison_VLC_RF}, VLC offers more flexibility in terms of the angle of incidence and beam-width with the same high accuracy percentage. In terms of physical size, the VLC transceiver is expected to be much smaller as it only needs a PD, which can be very small in size similar to the PDs used in \cite{Turan_2016_Physical_layer_Vehicular_Implementation} and \cite{Yeh_2013_Realtime_VLC_Compact_System}. On the other hand, RF systems need to have the transceiver module and the antenna which depends on the frequency of operation. In addition, some limitations that need to be considered in the future work are the channel model estimation in real-time and the performance during different light and environment scenarios. Due to the fact that light wave has a higher frequency than RF wave used in RF systems, the operation (distance) range in VLC will be smaller than that in RF, as expected. However,  VLC systems present immunity to EM interference (EMI), make use of unlicensed bands and present low power consumption (since VL is already used for illumination)  \cite{Uysal_2015_V2V_VLC_Channel,Cheng_2018_VLC_RF_CHannel_Comparison}. 
    
    \textbf{The motivating factors behind using VLC are summarized as follows:}

\begin{itemize}
        
        \item  \textbf{RF spectrum scarcity problem:} 
        
        The supporting band for the traditional RF communications is $300$ KHz - $300$ GHz. An application for wireless spectrum resources requires in general a high license fee and a long waiting period. With the rapid development of wireless communications services, the global wireless spectrum resources are in short supply. VL presents a huge bandwidth of 400-790 THz. Thus, the visible spectrum range is 10,000 times larger than that of the entire RF wireless spectrum. Furthermore, no license is required for VLC \cite{Perrig_2004_Security}.
        
        \item  \textbf{Reduced capacity of RF-based systems:}
        
        VLC technology assumes higher communication bandwidths by using higher frequency light waves to carry information. At the same time, due to the good directionality of the VL beam and weak diffraction, VLC can make use of diversity and multiplexing techniques to greatly expand the capacity of the communication system \cite{Jun_2003_Capacity}.
        
        \item  \textbf{Minimum EM interference:}
        
        VLC generates almost no EM radiation in space, which provides a good wireless communication solution for environments with EM interference sensitive devices or where radio silence is required \cite{Cheng_2018_VLC_RF_CHannel_Comparison}.
        
        \item  \textbf{Signal secrecy:}
        
        Due to the long wavelength and antenna architecture, RF communication presents weak directionality and can be easily tracked by non-target receivers. In VLC, since the light wavelength is short, light is subject to directionality and attenuation effects and cannot penetrate walls when it propagates in space. Thus,  light can only be received in a specific area and the confidentiality of transmitted information is ensured \cite{Mostafa_2014_Physical}. 
        
        \item  \textbf{Energy efficiency:}
        
        Compared to traditional light sources, LEDs present higher electro-optic conversion efficiency. If LEDs can be used to provide communication services while lighting, they will definitely save a lot of energy. Therefore, the concept of integrated lighting communication has also been proposed \cite{Din_2014_energy}. 
        Furthermore, compared to  RF femtocell networks, VLC systems could achieve a very large area spectral efficiency gain \cite{Stefan_2013_Area_spectral_efficiency_RF_VLC}.
        
        \item  \textbf{Health safety:}
        
        Compared with infrared radiation (IR) communications,  an improper application of IR can cause damage to the human body due to the high temperature since IR is a kind of heat radiation. For example, strong IR rays can cause skin burns or damage the retina of the fundus. Under normal lighting conditions, the use of VLC does not pose any safety problems for the human body \cite{Karunatilaka_2015_VLC_Survey}.
        
        \item  \textbf{Pervasive infrastructure (LEDs):}
        
        Due to the gradual widespread adoption of LEDs as the fourth generation of lighting technology, lighting facilities represent a natural platform for VLC. Only communication modules need to be added to the existing lighting facilities to implement the VLC function, so the installation cost is very low \cite{Fuada_2017_First}. Moreover, the current drives and controls the luminous intensity of LEDs, common VLC systems use the IM/DD scheme. When using LEDs for VLC, the interconnection of VL networks can be achieved through existing power line infrastructure \cite{Ayub_2013_Practical}. Therefore, VLC is compatible with the smart power grid.
        
        \item \textbf{Coherence time:}
        
        Since VLC is subject to significantly less multipath effects, the coherence time of the VLC channels is at least an order of magnitude larger than that of the RF channels. Thus,  VLC requires less frequent channel estimation, a feature which is especially important for situations that require continuous and stable linking. In addition, higher coherence time means that VLC is a good fault-tolerant technology \cite{Cheng_2018_VLC_RF_CHannel_Comparison}. 
        
    \end{itemize}
    \textbf{The key limiting factors of using VLC are summarized as follows:}    
    
    \begin{itemize}
        
       \item  \textbf{Uplink hardware issues:}
        Uplink communication for VLC represents a challenge as it requires updating the user's hardware. The signal emitted from the LED to the photodiode on the mobile phone only solves the downlink communication problem, and the mobile phone needs to send a signal back to ensure the communication link is unblocked, but the design of the reverse communication link is difficult \cite{Alresheedi_2017_Uplink}\footnote{Although the uplink in VLC is one of the main challenges, studies \cite{a_revision} have shown that hybrid VLC/IR (infrared) can be utilized to overcome this issue. Therefore, inherently a hybrid RF/VLC/IR system can be an ultimate solution for robust and high data rate indoor communication solution.}.

        \item  \textbf{Interference and noise from other light sources and inter-cell interference:}
        
         Other unmodulated artificial light sources and natural light sources sometimes work in the same spectral band with the VLC systems. If the ambient light source is strong enough, it will increase the intensity of the shot noise or saturate the receiving end and the VLC systems cannot function properly \cite{Cheng_2018_VLC_RF_CHannel_Comparison}. 
        
        \item   \textbf{Communication and lighting integration:}
        
        As a communication and lighting integration technology, indoor VLC must balance the dual user requirements of lighting and communication. The transmission of information in VLC relies on changing the luminous intensity, and flickering may occur during communication, which is not allowed in daily lighting and can seriously affect the user's lighting experience. VLC needs to meet the corresponding lighting requirements in both lightings on and lighting off modes \cite{Lou_2017_Joint}. 
        
        \item  \textbf{Terminal (user) mobility and handover overhead:} 
        
         If the receiver or transmitter is mobile, the received power on the detector array will fluctuate. The channel matrix will have to be updated over time.  Soft handover mechanisms are especially important to extract these fluctuations and maintain a more stable connection when the light detection is handed over from one photo-detector to another \cite{Vegni_2012_Handover}. 
        
        \item  \textbf{Signal coverage (the high attenuation rate of the signal):}
        
         In VLC, cell sizes are considerably smaller due to the high directivity of light and smaller transmission distances, thus the signal coverage area is limited. If the beam angle of the transmitter is increased, the detected signal intensity will be relatively reduced \cite{Vavoulas_2015_Coverage}. 
        
        \item  \textbf{Shadowing, due to losing the LOS link:} 
        
         As VL cannot penetrate obstacles, indoor VLC is built for the LOS link. Thus, the transmission is influenced by the blocking objects due to the random movement of people in a room. However, clear LOS is expected for the receivers of lighting systems at most time \cite{Xiang_2014_Human}.
        
        \item  \textbf{VLC network access to Internet:}
        
         The indoor VLC system must be connected to the base station to achieve the communication objective. The most practical problem is how to construct a VL wireless access network consisting of dozens or even hundreds of VL APs that are distributed over the ceiling since it is difficult to install new communication cables between different fixed networks and LED lights or among the LED lights \cite{Shao_2014_Indoor_Conf}.

	\end{itemize}
	
	Furthermore, more detailed discussion about the limitations of implementing VLC system is presented in section \ref{Sec:Hybrid_Network_Exp_implementation}.

			\begin{sidewaysfigure*}
		\includegraphics[scale = 0.39]{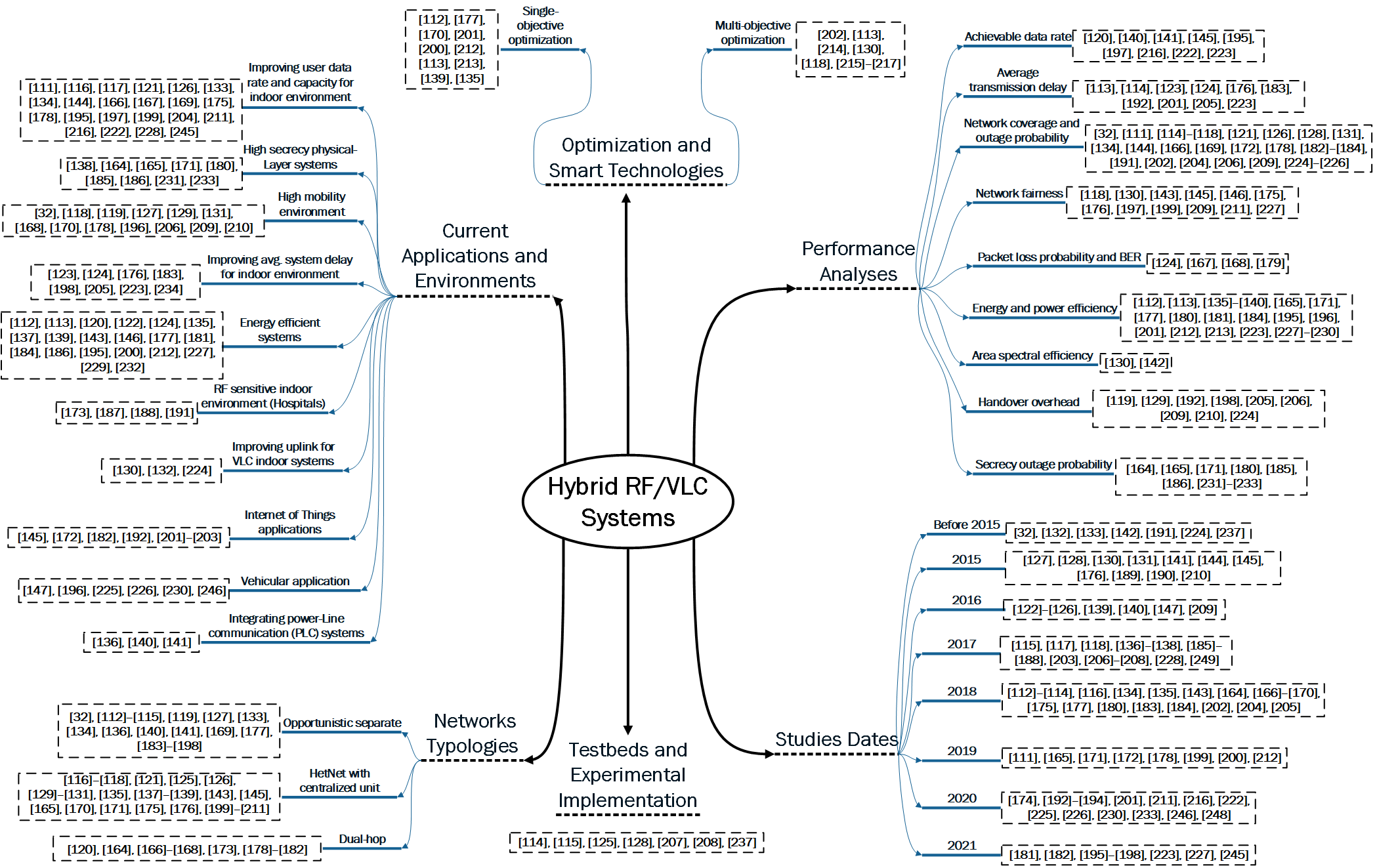}
		\centering
		
		\caption{Classification of hybrid RF/VLC systems.}
		\label{fig:Hybrid_Studies_tree}
	\end{sidewaysfigure*}

	\section{Hybrid systems}
	\label{sec:Hybrid_sys}
    This section discusses the hybrid RF/VLC studies in more detail. It is divided into the following subsections: hybrid networks typologies, hybrid RF/VLC environments, optimization and smart technologies, performance analyses, hybrid network simulation and system implementation, and current applications of hybrid systems.
    Fig.~\ref{fig:Hybrid_Studies_tree}  presents an overview of the main studies and it enables the readers to navigate easier through the topic of interest.
	
	\subsection{Hybrid Networks Topologies}
	
    The following three types of hybrid RF/VLC networks topologies were proposed in literature\footnote{While "Heterogeneous" is commonly used for  parallel hybrid RF/VLC networks, some references used it for dual-hop hybrid RF/VLC.}:
    
            \subsubsection{Dual-Hop Hybrid RF/VLC System} 
\begin{figure}[t!]
			\includegraphics[clip,trim=0 0 0 0, scale=0.26]{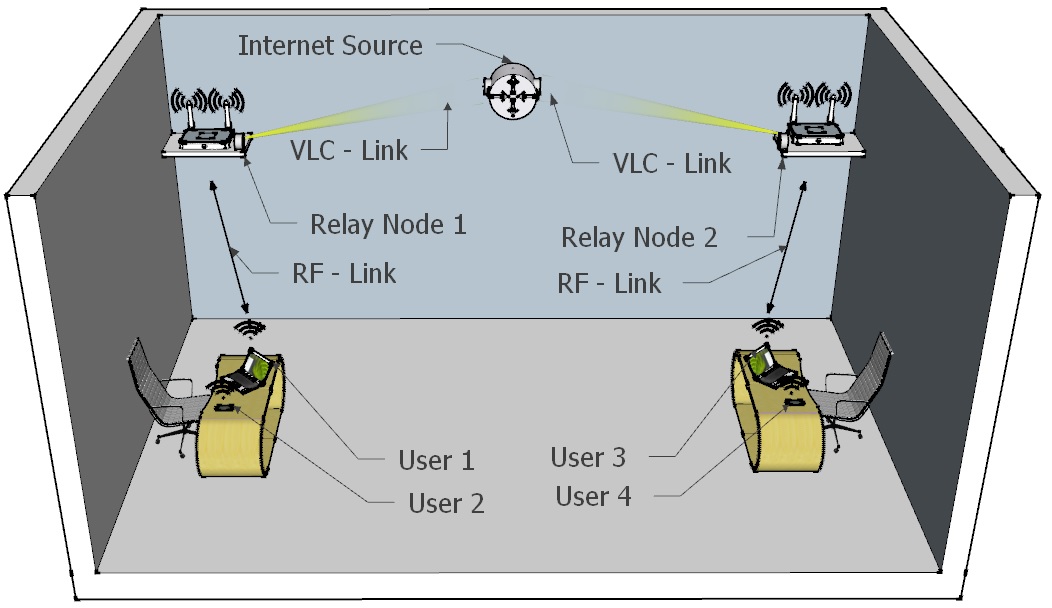}
			\centering
			\caption{Hybrid RF/VLC Dual-hop systems.}

			\label{fig:DualHop_sys_model}
		\end{figure}

In this type of hybrid system, the user access link is either VLC or RF. The network model where the user access link is RF can be observed in Fig.~\ref{fig:DualHop_sys_model}\footnote{Note that the VLC APs in Figs.~\ref{fig:DualHop_sys_model},~\ref{fig:HetNet(NoCentralUnit)}, and~\ref{fig:HetNet(WithCentralUnit)} are basically the ones that provides VLC links. Their placements would be application specific. These figures are provided for illustrative purposes. For example, the VLC APs are placed on the wall in Figure~\ref{fig:HetNet(NoCentralUnit)}. Different scenarios would be possible as well.}.

    The first component of the system is the communication between the VLC source (base station) and relaying nodes. The communication is carried out using VLC (LOS-link) in both uplink and downlink. Due to the fixed positions of the relay nodes and base-station, it is believed the maximum data rate can be achieved using the LOS VLC system. 
        
        The second component is the communication between the end-user and (plug-and-play) relay node (AP). The communication is accomplished using a low power directional RF system by dividing the area into cells using a  concept similar to that of mobile base stations. Dividing large areas into small cells will provide the advantage of serving multiple users and saving power, simultaneously. Furthermore, RF communication schemes such as handover, multi-relaying, and beamforming (via directional antennas) can be applied to this system, if needed. 
        
        In \cite{Liao_2018_Jour_Physical_Layer_Security}, the secrecy performance of a hybrid RF/VLC system was investigated assuming that the relay node extracts the direct current component and collects energy from the optical signal and then uses the collected energy for retransmitting data.  Exact and asymptotic expressions for secure outage probability and average secrecy capacity considering the effect of system parameters were also derived.  In \cite{Khori_2019_journal_Secrecy}, the secrecy capacity of DF-based hybrid RF/VLC was compared with standalone RF and VLC systems and it was showed that the hybrid had better performance. This paper also analyzed the framework of non-adaptive power allocation where both the source and relay present the same amount of power and the case of cooperative power-saving where the total average power is shared between source and relay in a way that minimizes the total power while satisfying the required secrecy capacity. In \cite{Zhang_2018_Conf_Coop_WSNs}, the effect of positions' randomness of both relay and destination on outage probability of cooperative hybrid RF/VLC wireless sensor networks was investigated. Decode \& forward (DF) and amplify \& forward (AF) relaying schemes were analyzed and approximate expressions for outage probability were derived. In \cite{Namdar_2018_jour_Outage_relay}, outage and BER performances of AF relay-assisted hybrid RF/VLC systems were investigated. Closed-form expressions for the outage probability were derived using probability density function (PDF) and moment generating function (MGF) approaches and by considering the effect of emission angle. The effect of timing errors on BER performance was also considered. In \cite{Zhang_2018_Jour_Spatially_Random}, taking into account the randomness of relay's and destination's locations, the outage and symbol error probabilities of a hybrid RF/VLC assuming DF and AF relaying schemes were derived. 
        
        In \cite{Han_2018_Conf_Bipartite_Matching}, a new multi-user hybrid RF/VLC was proposed. In this system,  users are divided into pairs. The near user receives from source via VLC and forwards the information to its paired user through RF transmission. In \cite{ Du_2018_Jour_Knowledge_Transfer}, a learning algorithm-knowledge transfer context-aware hybrid RF/VLC system that takes into account the traffic type, location, and time was proposed. The presented simulation results illustrated that the proposed system could significantly improve the convergence speed and performance of reinforcement learning-based network algorithms. In \cite{Khori_2019_Jour_Beamforming_Relaying}, the secrecy performance of relay-jammer selection beamforming hybrid RF/VLC was investigated assuming the absence of a direct link between source and destination. The considered system presents multiple DF relays and the relay node is selected by minimizing the outage probability. The jamming node is then selected from the available relaying nodes based on the received signal-to-noise ratio (SNR) at the eavesdropper location. Furthermore, beamforming vectors for both RF and VLC subsystems were designed and exploited in minimizing the consumed power. In \cite{Pan_2019_Jour_IoT}, the outage performance of the IoT hybrid RF/VLC system was investigated considering the randomness of the positions of devices.  VLC was considered for the downlink from the source lamp to the IoT devices while RF with the NOMA scheme for the uplink. All IoT devices are equipped with PD for two purposes: data communication and energy harvesting from the light emitted by the source LED lamp. These devices are then using the harvested energy to transmit data to the RF receiver. Approximate expressions for the outage probability were also derived.  In \cite{Vats_2017_Conf_Outage_E_health_AF}, a medical health care AF relaying RF/VLC system was proposed. In the proposed system, the RF link is for the outdoor link whereas the VLC is for the indoor link. The outage probability was also investigated assuming generalized K-fading in RF link.

		\subsubsection{Opportunistic Separate Networks (RF/VLC)}
		
\begin{figure}[t!]
			\includegraphics[clip,trim=0 0 0 5, scale=0.27]{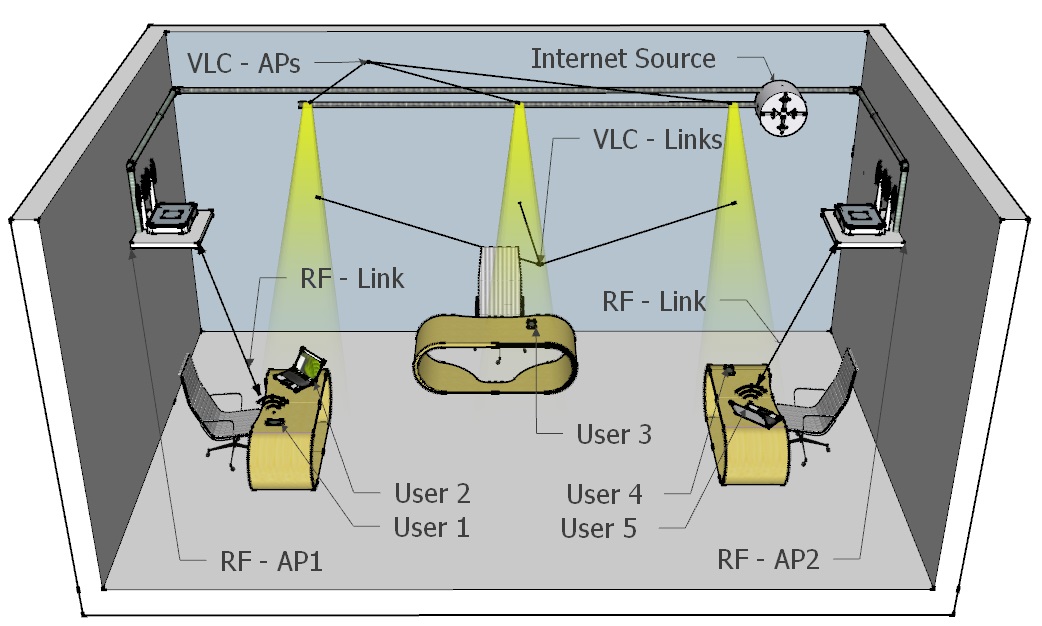}
			\centering
			
			\caption{Hybrid RF/VLC systems: Opportunistic separate networks.}
			\label{fig:HetNet(NoCentralUnit)}
		\end{figure}

        Two separate VLC and RF networks with the user deciding on the which system needs to be used. This network model is presented in Fig.~\ref{fig:HetNet(NoCentralUnit)}. In this network, the users are free to choose the best network depending on several parameters like SNR of the system, application requirements, mobility of the user, etc.
        This type of network is hard to control and optimize. Some of the studies adopted this model because of its low network overhead and its flexibility to the users. However, it is hard to control the users when there is a high number of users in the same place competing for the same resource without a control unit dividing the resources. Therefore, as the number of users increases, it is hard to adopt this model. 
        
        The coverage and rate analysis of opportunistic cellular RF/VLC and other network configurations were investigated and compared in \cite{Tabassum_2018_Coverage_Journal}. Based on approximations of the complementary error function (erfc) and cumulative distribution function (CDF) of a Gamma random variable, an approximate expression for coverage was derived. It was shown that the opportunistic selection based on the maximum received signal power is more suitable for scenarios where the interference effects are not dominant. On the other hand, the opportunistic scheme deteriorates the performance for higher interference scenarios, due to wrongful connection to RF networks with higher interference instead of VLC networks.
       Hybrid cellular RF/VLC has also been considered in \cite{Rajo_jornal_2020 }.  This paper presented both theory and implementation demonstrating the feasibility of heterogeneous RF/VLC network for IoT. Particularly, a hybrid cellular architecture that allows internet-operability between different technologies was proposed. They further discussed different applications such as localization, long-range communication, and monitoring.
        
        In \cite{Papanikolaou_2018_Conf_NOMA}, a hybrid RF/VLC system was investigated with one RF AP and multiple VLC APs assuming that all APs perform NOMA. Considering the fact that grouping users in the NOMA system is challenging, this reference addressed the grouping of users using the coalitional game theory. In particular,  a merge-and-split algorithm that determines the optimal user grouping is presented. A comparison of the proposed system with the conventional opportunistic scheme points out the effectiveness and robustness of the proposed scheme.
        Lastly, in opportunistic separate networks, all the nodes are competing on the resources. Therefore, the network optimization and control are challenging tasks, especially when there is no control unit handling the resource allocations between users.
	
		\subsubsection{Heterogeneous Networks (HetNet) with Centralized Unit}
		
\begin{figure}[t!]
			\includegraphics[clip,trim=0 0 0 0, scale=0.26]{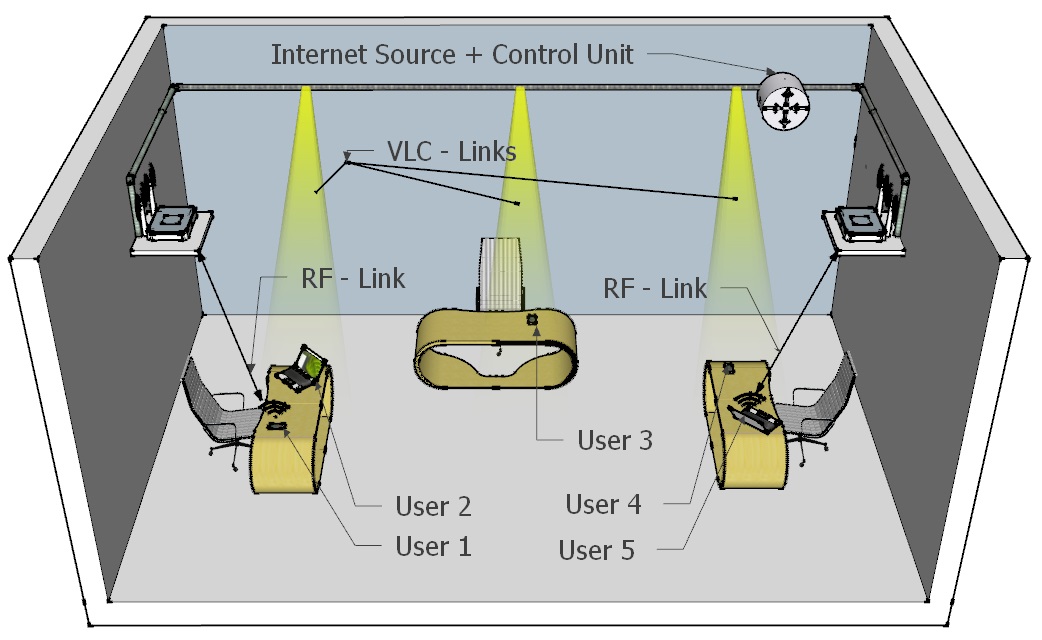}
			\centering
			\caption{Hybrid RF/VLC systems: Heterogeneous networks with centralized unit. }
			\label{fig:HetNet(WithCentralUnit)}
		\end{figure}

        As shown in Fig.~\ref{fig:HetNet(WithCentralUnit)}, the heterogeneous system consists of RF and VLC networks with a central control unit. Based on the network conditions, the central unit assigns the resources to the user, and sometimes having the location of the user helps the network to optimize the resources and interference better. 
        This type of network presents high network overhead data but it makes sure that all users are getting equal treatment (network fairness). Therefore, this network is needed when a high number of users are present in a small area network.

        In \cite{Jin_2015_Jour_Delay_Guarantee_Femtocell}, the optimal resource-allocation of mobile terminals in a heterogeneous wireless network under diverse quality of service (QoS) was considered. A decentralized algorithm was proposed to address the resource allocation problem. In \cite{Hsiao_2018_Conf_Ener_Eff_Max}, a heterogeneous cellular network that combines RF and VLC  to maximize the energy efficiency of the entire communication system under QoS requirements was considered. The optimization problem is not convex and it is addressed using successive convex approximations. In \cite{Duvnjak_2015_Heterogeneous}, a heterogeneous hybrid RF/VLC system where users can estimate their position based on the information broadcasted by VLC lamps was investigated. Based on the locations of users, the Wi-Fi unit allocates the resources of VLC enabled lamps. In \cite{Vats_2017_Conf_Outage_E_health_AF}, a medical health care AF relaying RF/VLC system was proposed. In the proposed system, the RF  is assumed for the outdoor link whereas the VLC is considered for the indoor link.

    Depending on system link-level topology of the system, the number of users and complexity of the system needed,  different authors chose the network model that fits their study as shown in Table~\ref{table:Comparison_Network_topology}.

	\begin{table}[t!]
		\footnotesize
		\centering
		\caption{Hybrid RF/VLC studies: Network topologies\label{table:Comparison_Network_topology}.}
		\renewcommand{\arraystretch}{1.1}
		\begin{tabular}{|>{\centering\arraybackslash}m{3cm}|>{\centering\arraybackslash}m{4.5cm}|}
			\hline
			\textbf{Network topology}                                              & \textbf{References}
		\end{tabular}
		\begin{tabular}{|>{\centering\arraybackslash}m{3cm}|>{\arraybackslash}m{4.5cm}|}
			\hline
			Dual-hop  & $\!\!\!$\cite{Chowdhury_2019_Integrated_Journal,Petkovic_2019_Mixed_Conf,Zenaidi_2017_Achievable_Conf, Zhang_2018_Jour_Spatially_Random, khori_2018_Conf_Phys_Lay_Secu, Namdar_2018_jour_Outage_relay, Zhang_2018_Conf_Coop_WSNs, Liao_2018_Jour_Physical_Layer_Security, Vats_2017_Conf_Outage_E_health_AF,Xiao_Cooperative_2021_Journal,Peng_EndtoEnd_2021_Journal} \\\hline
			
			Opportunistic  separate & $\!\!\!$\cite{Tabassum_2018_Coverage_Journal, Wu_2018_Jour_adap_resou_opti, Han_2018_Conf_Bipartite_Matching, Hammouda_2018_Conf_QoS_Constr, Khreishah_2018_jour_Energy_Efficient, Li_2018_Conf_BW_Aggreg, Zhou_2018_Conf_Coop_NOMA, Hsiao_2018_Conf_Ener_Eff_Max, Kashef_2017_Transmit_PLC_Journal, Pan_2017_Secrecy_Conf,Pan_2017_Secure_Journal, Vats_2017_Hybrid_Heath_Journal,Vats_2017_Modeling_Heath_Conf, Wentao_2017_Design_Journal, Bao_2017_Visible_HetNet_Journal, Kashef_2015_Achievable_Conf,Kashef_2016_Impact_PLC_Conf, Rahaim_2015_Toward_Journal, Little_2015_Network_topologies_Conf, Duvnjak_2015_Heterogeneous, Chowdhury_2014_Cooperative_Journal, Hussain_2014_Hybrid_Conf, Rahaim_2011_hybrid_Conf,Wu_Data_Driven_2020_Journal,Alenezi_Reinforcement_2020_Conf,Adnan_Load_Balancing_2020_Journal,Hammadi_NonOrthogonal_2021_Journal,Chen_Coordination_2021_Journal,Aboagye_Joint_2021_Journal,Arshad_Stochastic_2021_Journal} \\\hline
			
			HetNet with centralized unit & $\!\!\!$\cite{Papanikolaou_2019_User_NOMA_Journal,Tran_2019_Ultra_Jorunal,Yang_2020_Learning_IoT_Journal,Khori_2019_journal_Secrecy, Khori_2019_Jour_Beamforming_Relaying, Becvar_2018_Conf_D2D, Mach_2017_Conf_D2D, Zhang_2018_Energy_Journal, Obeed_2018_Jour_power_allocation, Du_2018_Jour_Knowledge_Transfer, Pratama_2018_Jour_BW_Aggreg, Papanikolaou_2018_Conf_NOMA, Bao_2018_Conf_QoE_VHO, Wang_2017_Optimization_Journal, Wang_2018_Learning_Journal, Wu_2017_Joint_Conf, Wu_2016_Two_stage_Conf,Wu_2017_Access_Journal, Marzban_2017_Beamforming_Conf, Wu_2017_Access_Conf, Kafafy_2017_Power_Conf, Saud_2017_Conf_Software_Defined,Saud_2017_Conf_Software_Defined2, Kashef_2016_Energy_Journal, Li_2016_Mobility_Letter, Yan_2016_combination_Conf, Li_2015_Cooperative_Journal, Wang_2015_Dynamic_Journal, Jin_2015_Jour_Delay_Guarantee_Femtocell, Wang_2015_Efficient_Journal, Wang_2015_Distributed_Conf, Liang_2015_Conf_LTE,Amjad_Optimization_2020_Journal}\\\hline

		\end{tabular}
	\end{table}


    In terms of the utilization of the system resources, there are two categories that can describe a hybrid RF/VLC system:
    \begin{itemize}
        \item \textbf{Aggregated Technology:}
        
        Where users employ both VLC and RF technologies at the same time to improve their data rate and reliability of the connection.
        
        \item \textbf{Non-Aggregated Technology:}
        
        Where users employ only one of the two technologies to optimize the network conditions and manage the interference present in the network.
    \end{itemize}

\begin{table*}[t]
	\footnotesize
		\centering
		\caption{Hybrid RF/VLC studies: Objective optimization problems.}
		\label{table:Optimization_Table}
		\linespread{1.15}\selectfont   

		\begin{tabular}{|>{\centering\arraybackslash}m{1.2cm}|>{\centering\arraybackslash}m{13cm}|}
			\hline
			\multicolumn{2}{|c|}{\textbf{Single-objective optimization problem}} \\\hline
			\textbf{References}  & \textbf{Objective}   \\ \hline	
	\end{tabular}\\
	\begin{tabular}{|>{\centering\arraybackslash}m{1.2cm}|>{\arraybackslash}m{13cm}|}
			\cite{Khreishah_2018_jour_Energy_Efficient} & Minimizing the power consumption subject to satisfying the users' requests and maintaining an acceptable illumination level.\\\hline
			{\cite{Hsiao_2018_Conf_Ener_Eff_Max}} &  Maximizing energy efficiency of the whole
communication system under QoS requirements.\\\hline
			\cite{ Du_2018_Jour_Knowledge_Transfer}& Selecting the network using machine learning methods, the network that provides the best long term average performance assuming all participants have learning abilities. \\\hline 
			\cite{Yang_2020_Learning_IoT_Journal} & Maximizing the network energy efficiency by jointly considering network selection, channel allocation, and power management and formulated the decision-making problem as a MDP.\\ \hline
			\cite{ Tran_2019_Ultra_Jorunal} & Enhancing the energy harvesting of the overall network that consist of one multi-antenna RF AP, multiple optical transmitters, and multiple terminal devices.\\ \hline
			\cite{Kong_2019_Energy_Jorunal } & Minimizing the area power consumption (APC) under an outage probability constraint. \\ \hline
		    \cite{Ma_Location_Conf} & Maximizing the throughput under proportional fairness constraints. \\ \hline
			\cite{Amjad_2019_Optimization_Conf} & Allocation of available APs with the objective of increasing the overall throughput.\\ \hline
			\cite{Kashef_2016_Energy_Journal} & Optimize energy efficiency of heterogeneous RF/VLC network that has a single RF AP and multiple VLC APs. \\\hline
			\cite{Zhang_2018_Energy_Journal} & Optimizing energy efficiency of heterogeneous RF/VLC network that has a multiple RF APs and multiple VLC APs while meeting devices' QoS. \\\hline
		\end{tabular}\\
	\begin{tabular}{|>{\centering\arraybackslash}m{1.2cm}|>{\centering\arraybackslash}m{13cm}|}
			\multicolumn{2}{|c|}{\textbf{Multi-objective optimization problem}} \\ \hline
			\textbf{References}  & \textbf{Objective}   \\ \hline
	\end{tabular}\\
	\begin{tabular}{|>{\centering\arraybackslash}m{1.2cm}|>{\arraybackslash}m{13cm}|>{\arraybackslash}m{5.5cm}|}
		 \cite{Becvar_2018_Conf_D2D}  & Outage reduction and system capacity improvement. \\\hline
		 \cite{Wu_2018_Jour_adap_resou_opti} & Reducing communications power consumption and minimizing the queue lengths.\\\hline
		  \cite{Aboagye_ICC_Optimization} & Jointly optimizing user association and power control considering the required data rates and the maximum available transmit power for RF base stations (BSs) and the VLC in three-tier HetNet.\\\hline
		 \cite{Li_2015_Cooperative_Journal} & Providing a higher Mean Bandwidth Efficiency of VLC network and higher average throughput for the hybrid RF/VLC by employing Vectored Transmission  Techniques. \\\hline
         \cite{Wang_2017_Optimization_Journal} & Jointly and separately optimizing AP assignment and resource allocation considering the effect of users' mobility in an effort to increase the network's capacity.  \\\hline
		 \cite{Wang_Journal_Opt_2017} & Achieving better user satisfaction performance with low computational complexity by jointly dealing with AP assignment and resource allocation using Evolutionary Game Theory. \\\hline
		 \cite{Zeng_Realistic_2020_Journal} & This is an extension of \cite{Wang_Journal_Opt_2017}  where  the orientation-based random waypoint mobility model has been considered with link selection and resource allocation optimization problem. \\\hline
		  \cite{Chen_TWC_opti_2020} & Adapting transmit power of both RF and VLC considering the tradeoff between power consumption and delay-limited connectivity where the topology control problem is modeled as a potential game.\\\hline

		\end{tabular}
\end{table*}

   \subsection{Resource Allocation}
    Different hybrid RF/VLC optimization problems spanning different topics such as resource allocation, transmit power minimization, load balancing and handover have been considered in the literature and are summarized in Table~\ref{table:Optimization_Table}. These optimization problems can be divided generally into two categories:

    \subsubsection{Single-Objective Optimization (SOO)} SOO is an optimization problem with a single objective function. The optimization problem of resource allocation where the heterogeneous Hybrid RF/VLC network has a single RF AP and multiple VLC APs was covered in \cite{Kashef_2016_Energy_Journal,Kong_2019_Energy_Jorunal}, with the objective of optimizing energy efficiency while meeting devices' QoSs. As a larger system with similar objective of energy efficient Hybrid VLC/RF optimization problem, several RF APs and VLC APs have been considered in\cite{Zhang_2018_Energy_Journal}. In \cite{Khreishah_2018_jour_Energy_Efficient}, authors consider an energy-efficient hybrid RF/VLC system. The objective of the optimization problem is to minimize the power consumption subject to satisfying the users’ requests and maintaining an acceptable illumination level. In \cite{ Du_2018_Jour_Knowledge_Transfer}, they have considered the problem of selecting the network that provides the best long-term average performance where the network selection method is based on machine learning that assumes all participants have learning/cognitive abilities. Authors in \cite{Hsiao_2018_Conf_Ener_Eff_Max} investigate a heterogeneous cellular network that combines RF and VLC in order to maximize the energy efficiency of the whole communication system under QoS requirements. In \cite{Yang_2020_Learning_IoT_Journal }, a heterogeneous hybrid RF/VLC for wireless industrial networks is developed to support different QoS requirements such as high reliability, low latency and high data rates of IoT and industrial networks IoT (IIoT) devices. They have proposed a new deep post-decision state-based experience replay and transfer (PDS-ERT) learning algorithm in order to maximize the network energy efficiency. Particularly, they have considered an energy-efficient resource management problem by jointly considering network selection, channel allocation, and power management and formulated the decision-making problem as a Markov decision process (MDP). They have then proposed a new PDS-ERT learning algorithm to learn the optimal policy, which boost up the learning speed and enhance the learning efficiency. On the other hand, in \cite{Tran_2019_Ultra_Jorunal} a collaborative RF with lightwave resource allocation scheme was proposed for enhancing the energy harvesting of the overall network which consists of one multi-antenna RF AP, multiple optical transmitters, and multiple terminal devices. Under an outage probability constraint, minimizing the area power consumption (APC) by optimizing the RF and VLC BSs intensities in \cite{Kong_2019_Energy_Jorunal }. Utilizing the knowledge of location information of user terminals, a location information-aided load balancing (LB) design for the hybrid RF/VLC networks was proposed in \cite{Ma_Location_Conf} to maximize the throughput under proportional fairness constraints.
    In \cite{Kong_2019_Energy_Jorunal }, authors have designed energy efficient hybrid RF/VLC networks by considering a constraint on the outage probability. They have, particularly, proposed new strategies that optimize the intensities of macro cell BSs (MBSs), small cell BSs (SBSs) and VLC BSs (VLCBSs). They have left the energy consumption caused by handover  for future studies. In \cite{Amjad_2019_Optimization_Conf}, a hybrid RF/VLC network that consists of more than one RF and VLC APs has been considered. Therefore, they have increased the degree of freedom for selecting the AP that the node should be connected to. Thus, an optimization problem for the allocation of these APs with the objective of increasing the overall throughput was proposed.
    
    \subsubsection{Multi-Objective Optimization (MOO)} MOO is an optimization problem that considers more than one objective function simultaneously or sequentially (i.e., adopts multiple objective functions). Several studies have concentrated on load balancing optimization problem in heterogeneous RF/VLC networks in order to increase the network's capacity and improve devices fairness \cite{Li_2015_Cooperative_Journal,Wang_2017_Optimization_Journal}. In \cite{Wang_2017_Optimization_Journal}, a dynamic load balancing scheme for Hybrid RF/VLC networks is proposed considering the effect of users' mobility. They have considered both joint and separate optimization algorithms which are, respectively, jointly and separately optimize AP assignment and resource allocation. In \cite{Li_2015_Cooperative_Journal}, LB problem in the context of hybrid RF/VLC systems has been considered. They have employed vectored transmission (VT) techniques among all VLC APs. As a result, VLC network becomes capable of providing a higher Mean Bandwidth Efficiency (MBE) and the hybrid RF/VLC system become capable of providing a higher average throughput. When the link selection and resource allocation optimization problem in heterogeneous RF/VLC  is classified as non-deterministic polynomial (NP), the coalitional game theory is used \cite{Wang_Journal_Opt_2017,Zeng_Realistic_2020_Journal, Chen_TWC_opti_2020}. In \cite{Wang_Journal_Opt_2017}, EGT based LB for an indoor hybrid RF/Li-Fi network has been considered. This system jointly deals with the AP assignment and resource allocation. They have demonstrated that their proposed EGT-based LB scheme is able to achieve a better user satisfaction performance with low computational complexity. They have then extended their work in \cite{Zeng_Realistic_2020_Journal}. In \cite{Zeng_Realistic_2020_Journal}, The orientation-based random waypoint mobility model has been considered in a hybrid Li-Fi/WiFi network to support dynamic load balancing for mobile users. In \cite{Chen_TWC_opti_2020}, RF and VLC transmitting power of each vehicle in Hybrid RF/VLC vehicular ad-hoc network is adapted locally based on the vehicle's knowledge about the vehicular ad-hoc network topology considering the tradeoff between power consumption and delay-limited connectivity. In \cite{Becvar_2018_Conf_D2D}, multi-objective optimization problem for selecting either RF or VLC at each device-to-device pair in a multi-user communication network has been investigated. The proposed solution consists of a two-phase algorithm in which the change from RF to VLC depends on the interference caused to other devices and the interference received from other device-to-device pairs. The first phase of the algorithm is for outage probability reduction while the second phase is for system capacity improvement. It is shown that the performance of the proposed algorithm is close to the exhaustive search algorithm that minimizes outage and maximizes capacity but presents less computational complexity. Reference  \cite{Wu_2018_Jour_adap_resou_opti} investigates the optimization of network resources in hybrid RF/VLC networks to reduce energy consumption for the communication task while minimizing the queue lengths.    In \cite{Aboagye_ICC_Optimization}, they have considered three-tier HetNet by introducing VLC into a two-tier HetNet and investigated the performance in terms of system's throughput, energy efficiency (EE), and spectral efficiency (SE).  The optimization problem has been formulated to jointly optimize user association and power control considering the required data rates and the maximum available transmit power for RF BSs and the VLC.

    
    In optimizing hybrid RF/VLC communication systems, an important problem is establishing the constraint conditions. Since the hybrid communication systems include both RF and VLC transceivers, the constraints are divided into two categories: lighting and coupling constraints, respectively.
    \begin{itemize}
        \item {\textbf{Lighting Constraints:}} 
        
        Since VLC needs to maintain illumination and communication requirements simultaneously, several constraints should be considered. Due to the requirement of intensity modulation in VLC, the transmit signal has to satisfy the non-negative constraint \cite{Gao_2016_Modulation}. In order to avoid excessive nonlinear distortions, it is sufficient to consider the peak optical power or a peak-to-average-power ratio (PAPR), which in general are constrained to guarantee that the LED is working in the linear region \cite{Gong_2015_Power}. The transmission of information in VLC relies on changing the luminous intensity. However,  flickering may occur during communication, a condition that is not allowed in daily lighting and can seriously affect the user's lighting experience. For illumination purposes, dimming constraints are desirable features to consider \cite{Gancarz_2013_Impact}. For multi-color VLC, it is necessary to ensure that the illumination meets the white light constraint. Using MacAdam ellipse as a statistical measuring tool, the small chromaticity difference between two colors in the same luminance color map can be described. Due to the limitation of human eyes in color recognition, when two colors are on the same MacAdam ellipse, ordinary human observers cannot distinguish between the two colors \cite{Gong_2015_Power}.

        \item {\textbf{Coupling Constraints:}}
        
        In the dual-hop RF/VLC systems, the RF and VLC systems work in serial, the data rate of the mobile terminals are constrained by the system where the transmission speed is slower \cite{Rakia_2016_Optimal_DualHop_Journal}. In heterogeneous RF/VLC systems, the RF and VLC systems work in parallel such that the mobile terminals aggregate the data from both networks. Hence, the data rate at the mobile terminals is constrained by the total data rate of the RF and VLC systems \cite{Kashef_2016_Impact_PLC_Conf}. The optimization problem is constrained by the power budget for these two serial/parallel communication links such that $P_\text{t-VLC}+P_\text{t-RF} \le P_\text{t}$, where $P_\text{t}$ is the maximum allowable total power\cite{Kashef_2015_Achievable_Conf}. In \cite{Liu_2018_Correlation}, the channel correlation in VLC over broad spectra is analyzed. However,  for hybrid RF/VLC systems, the coupling constraints of the RF and VLC channels have not been investigated yet and it represents an important problem for future studies. 
    \end{itemize}


	\subsection{Performance Analyses}
	Most of the existing studies have conducted different performance analyses. In Table~\ref{table:Comparison_Perfromance_analysis}, the studies discussing the same investigations  are grouped to enable  researchers an easy comparison. Moreover, a list of the main studies carried out is presented next.

				\begin{table*}
	\footnotesize
		\begin{center}
		\caption{Hybrid RF/VLC studies: Performance analyses.}
			\label{table:Comparison_Perfromance_analysis}
				\begin{tabular}{|>{\centering\arraybackslash}m{5.0cm}|>{\centering\arraybackslash}m{7.0cm}|}
					\hline
					\textbf{Analysis} & \textbf{References} \\
				\end{tabular}
				\begin{tabular}{|>{\arraybackslash}m{5cm}|>{\arraybackslash}m{7.0cm}|}
					\hline 
					Achievable data-rate & $\!\!$\cite{Zenaidi_2017_Achievable_Conf,Wang_2015_Distributed_Conf,Kashef_2015_Achievable_Conf,Kashef_2016_Impact_PLC_Conf,Zeng_Realistic_2020_Journal,Papanikolaou_Optimal_2020_Journal,Hammadi_NonOrthogonal_2021_Journal,Aboagye_Joint_2021_Journal,KUCUK_selfadaptive_2021_Journal}  \\  \hline 
					Average transmission delay & $\!\!\!$\cite{Yang_2020_Learning_IoT_Journal,Bao_2018_Conf_QoE_VHO,Hammouda_2018_Conf_QoS_Constr,Li_2018_Conf_BW_Aggreg,Wu_2018_Jour_adap_resou_opti,Jin_2015_Jour_Delay_Guarantee_Femtocell,Shao_2016_Delay_Journal,Rakia_2016_DualHop_VLC_RF_Conf,Wu_Data_Driven_2020_Journal,KUCUK_selfadaptive_2021_Journal}  \\ \hline
					Packet loss probability and BER & $\!\!\!$\cite{Petkovic_2019_Mixed_Conf,Rakia_2016_DualHop_VLC_RF_Conf,Namdar_2018_jour_Outage_relay,Zhang_2018_Jour_Spatially_Random} \\ \hline
					Network coverage and outage probability  & $\!\!$\cite{Chowdhury_2019_Integrated_Journal,Ma_Location_Conf,Becvar_2018_Conf_D2D,Pan_2019_Jour_IoT,Pratama_2018_Jour_BW_Aggreg,Han_2018_Conf_Bipartite_Matching,Zhou_2018_Conf_Coop_NOMA,Hammouda_2018_Conf_QoS_Constr,Li_2018_Conf_BW_Aggreg,Chowdhury_2014_Cooperative_Journal,Bao_2014_Protocol_Journal,Wang_2018_Learning_Journal,Wu_2017_Joint_Conf,Hussain_2014_Hybrid_Conf,Shao_2015_Design_Journal,Wu_2017_Access_Conf,Basnayaka_2015_Hybrid_Conf,Wu_2016_Two_stage_Conf,Wu_2017_Access_Journal,Li_2016_Mobility_Letter,Wang_2015_Dynamic_Journal,Wang_2017_Optimization_Journal,Zhang_2018_Conf_Coop_WSNs,Tabassum_2018_Coverage_Journal,Wentao_2017_Design_Journal,Nauryzbayev_Outage_2020_Journal,Abouzohri_Performance_2020_Conf,Peng_EndtoEnd_2021_Journal}  \\ \hline
					 Network fairness & $\!\!\!$\cite{Li_2016_Mobility_Letter,Li_2015_Cooperative_Journal,Obeed_2017_Joint_Conf,Wang_2015_Distributed_Conf,Wang_2017_Optimization_Journal,Papanikolaou_2018_Conf_NOMA,Papanikolaou_2019_User_NOMA_Journal,Jin_2015_Jour_Delay_Guarantee_Femtocell,Obeed_2018_Jour_power_allocation,Amjad_Optimization_2020_Journal,Obeed_Power_2021_letter,Aboagye_Joint_2021_Journal}  \\ \hline
					 Handover overhead  & $\!\!$\cite{Li_2016_Mobility_Letter,Wu_2017_Joint_Conf,Bao_2017_Visible_HetNet_Journal,Wang_2015_Efficient_Journal,Bao_2018_Conf_QoE_VHO,Liang_2015_Conf_LTE,Bao_2014_Protocol_Journal,Wu_Data_Driven_2020_Journal,Arshad_Stochastic_2021_Journal}  \\ \hline
					Energy and power efficiency & $\!\!\!$\cite{Amjad_2019_Optimization_Conf,Kong_2019_Energy_Jorunal,Yang_2020_Learning_IoT_Journal,Khori_2019_journal_Secrecy,Khori_2019_Jour_Beamforming_Relaying,Zhou_2018_Conf_Coop_NOMA,khori_2018_Conf_Phys_Lay_Secu,Khreishah_2018_jour_Energy_Efficient,Hsiao_2018_Conf_Ener_Eff_Max,Wu_2018_Jour_adap_resou_opti,Ashimbayeva_2017_Conf_Hard_Soft_Switching,Zhang_2018_Energy_Journal,Kashef_2017_Transmit_PLC_Journal,Kafafy_2017_Power_Conf,Kashef_2016_Energy_Journal,Marzban_2017_Beamforming_Conf,Wu_2017_Dynamic_Conf,Kashef_2016_Impact_PLC_Conf,Chen_Resource_2020_Letter,Xiao_Cooperative_2021_Journal,Hammadi_NonOrthogonal_2021_Journal,Chen_Coordination_2021_Journal,Obeed_Power_2021_letter,KUCUK_selfadaptive_2021_Journal}  \\ \hline
					Area spectral efficiency & $\!\!$\cite{Li_2015_Cooperative_Journal,Stefan_2014_Hybrid_Conf}  \\ \hline
					Secrecy outage probability and secrecy rate & $\!\!\!$\cite{Khori_2019_Conf_Secrecy_Two_Stages,Khori_2019_journal_Secrecy,Khori_2019_Jour_Beamforming_Relaying,khori_2018_Conf_Phys_Lay_Secu,Pan_2017_3DHybrid_Conf,Pan_2017_Secrecy_Conf,Pan_2017_Secure_Journal,Liao_2018_Jour_Physical_Layer_Security,Kumar_PLS_Analysis_2020_Journal}    \\ \hline
				\end{tabular}
		\end{center}
	\end{table*}
	

	\begin{itemize}

			\item \textbf{User throughput (achievable data rate) of RF/VLC networks:}
			
			The average achievable data rate is a critical performance metric in any wireless network used to estimate the average user throughput in the network given a certain bandwidth. The average data rate of a user ($R_{i}$) can be expressed as follows:
			\begin{equation}
		    \label{eq:Avg_data_rate}
	        R_{i} = {E}\left[B\,\log_{2}\left(1+ \gamma_i\left(t\right)\right)\right],
	    	\end{equation}
	    	where ${E}\left[\cdot\right]$ is the expected value operator with respect to time domain, $\gamma_i(t)$ is the signal to interference and noise ratio which changes over time ($t$) depending on the user location and technology used (VLC or RF) since each technology has different channel model and $B$ is the user assigned bandwidth.

	        In \cite{Wu_2016_Two_stage_Conf} and \cite{Wu_2017_Access_Journal}, an efficient AP selection scheme for hybrid RF/VLC networks is proposed. Based on the different channel characteristics,  a tailor-made AP selection method for the hybrid network is formulated as a two-stage algorithm, which first determines the users that need service from Wi-Fi and then performs AP selection for the remaining users. The simulation results show that the proposed method achieves a near-optimal throughput.
            
            Reference \cite{Wentao_2017_Design_Journal} proposed an indoor VLC-Wi-Fi hybrid network experimental platform that integrates multiple links with multiple access, hybrid network protocol, user mobility management mechanisms, and cell handover. The proposed hybrid VLC-Wi-Fi network exhibits better coverage and greater network capacity.
            In \cite{Wang_2015_Dynamic_Journal}, a dynamic load balancing scheme that considers handover overhead in a hybrid RF/VLC network is proposed.  The throughput performance is analyzed across the service areas and the effects of the handover overhead on handover locations and user throughput are discussed.

			\item \textbf{Average transmission delay (end-to-end delay):}

		   Average transmission delay is the average time needed to send a packet from source (transmitter) to the receiver in the network. This average time includes all the multi-hop retransmission time ($T^{i}_{tr}$) (time needed by middle nodes to process and retransmit the data), error retransmission time ($T_{err}$) (time needed when a new data packet is retransmitted due to an error). Moreover, this parameter includes the time delay added due to the handover process ($T_{HO}$).
		   The average transmission delay ($T_{avg}$) can be expressed as follows:
			\begin{equation}
		    \label{eq:Avg_trans_delay}
		    \begin{split}
{T_{avg}} = E\left[ \begin{array}{l}
\sum\limits_{i = 1}^N {\left( {T_{tr}^i\left( t \right)} \right)}  + {T_{err}}\left( t \right){{P}_{err}}\\
 \quad\;\;+ {T_{HO}}\left( t \right){{P}_{HO}}
\end{array} \right],
            \end{split}
	    	\end{equation}   
	    	
	    	\noindent where ${P}_{HO}$ and ${P}_{err}$ are, respectively, the probability that a handover process is needed and a transmission error occurred. In (4), $N$ is the maximum number of hops in the system.

			In \cite{Rakia_2016_Optimal_DualHop_Journal} and \cite{Rakia_2016_DualHop_VLC_RF_Conf}, the relay between the two hops in a  dual-hop RF/VLC transmission system is used to energy harvest from different artificial light sources and sunlight. The total time to transmit one data packet from the source to the receiver must satisfy a strict delay constraint. The statistical model for the harvested energy at the relay is further proposed to analyze the packet loss probability.
			In \cite{Shao_2014_Indoor_Conf,Shao_2016_Delay_Journal}, the delay analysis of unsaturated heterogeneous omnidirectional/directional small cell hybrid RF/VLC wireless networks was investigated. In the first case, the minimum average system delay of the aggregated scenario is always lower than that of the non-aggregated scenario. In the second case, the heterogeneous contention-based omnidirectional small cells - directional small cell (CBOSC-DSC) network is studied. Extensive simulation results illustrate that the non-aggregated scenario outperforms the aggregated scenario due to the overhead caused by contention. In \cite{Hammouda_2018_Link_Journal}, the handover delay which would occur when the transmitter moves from one link to the other was investigated. In addition, the non-asymptotic bounds on data buffering delay were derived as well.

		    \item \textbf{Packet loss probability and bit error rate (BER):}
			
			BER is a ratio of the number of bit errors to the number of bits sent during a certain time interval. Since BER is a ratio of two numbers, it is a unitless parameter and sometimes expressed as a percentage. Moreover, packet loss probability can be expressed in terms of BER for a packet size of $L$ bits packet error rate (PER) is $BER\times L$. Finally, the error rate depends heavily on channel conditions, signal-to-interference-plus-noise ratio (SINR), error coding rate, and modulation type.

			In \cite{Rakia_2016_DualHop_VLC_RF_Conf}, the packet loss probability is analyzed using a  statistical model for the harvested electrical power and the time dedicated for excess energy harvesting. The data packet retransmission rate presents an optimal value that minimizes the packet loss probability, is independent of the RF channel path loss, and is inversely dependent on the packet size. In \cite{Namdar_2018_jour_Outage_relay}, outage and BER performances of AF relay-assisted hybrid RF/VLC system have been investigated.  Closed-form expressions for the outage probability are derived by considering the effect of emission angle. The effect of timing errors on BER performance has also been considered. In \cite{Zhang_2018_Jour_Spatially_Random}, considering the randomness of both relay's and destination's locations, the outage and symbol error probabilities of  hybrid RF/VLC with DF and AF relaying schemes have been studied.

			\item \textbf{Network coverage and outage probability:}
			
			The signal coverage probability ($P_c$) for an average user is defined as the probability that the users' instantaneous SINR ($\gamma_{o}$) is higher than a target SINR threshold ($\gamma_{th}$) given as:
			\begin{equation}
		    \label{eq:Signal_coverage_Prob}
	        P_c = {P}\left[\gamma_{o}>\gamma_{th}\right].
	    	\end{equation}
	    	
	    	Usually, RF networks have better coverage than VLC networks due to the difference in the wave propagation in both technologies. On the other hand, the outage probability ($P_{out}$) is the probability that the users' data rate is less than a threshold minimum data rate which is effectively the inverse of $P_c$.

            In \cite{Zhang_2018_Conf_Coop_WSNs}, the effect of positions' randomness of both relay and destination nodes on the outage probability of cooperative hybrid RF/VLC wireless sensor networks was investigated. Both DF and AF relaying schemes were considered and approximate expressions for outage probability were proposed and verified via Monte Carlo simulations. 
            In \cite{Pan_2019_Jour_IoT}, the outage performance of an IoT hybrid RF/VLC system was investigated by considering the randomness of the positions of devices. VLC was considered for the downlink from the source lamp to the IoT devices and RF with NOMA scheme for the uplink. All IoT devices are equipped with PD for two purposes: data communication and energy harvesting from the light emitted by the source LED lamp. These devices are then using the harvested energy to transmit data to the RF receiver.  Approximate expressions for the outage probability were also derived and validated via Monte Carlo simulations.
        
            In \cite{Yan_2016_combination_Conf}, a hybrid RF/VLC based indoor wireless access network structure is proposed. In this paper, the relationship between the wireless signal quality and the distance in a typical family or small business indoor layout is analyzed to develop the handover scheme.
            Reference \cite{Shao_2015_Design_Journal} assessed the throughput across various horizontal distances within the coverage of the VLC source and the results showed that the aggregating systems can achieve higher throughput. In \cite{Tabassum_2018_Coverage_Journal},  a unified framework is presented for the coverage and rate analysis of coexisting cellular RF/VLC networks under different network configurations.

			\item \textbf{Network fairness and users' satisfaction:}
			
			CDF of users' satisfaction for various numbers of users is calculated and used to measure the fairness of the network among the users. Fairness parameter is used to determine if users are equally utilizing the network resources. There are multiple ways to measure the network fairness, one of the most famous metrics is Jain's fairness index ($F$) which is calculated as follows \cite{Obeed_2018_Jour_power_allocation}:
			\begin{equation}
		    \label{eq:Network_Fairness}
	        F = \frac{(\sum_{i=1}^{N_{AP}}R_{i})^2 }{(N_{AP})\sum_{i=1}^{N_{AP}}(R_{i})^2},
	    	\end{equation}
	    	where $N_{AP}$ is the number of APs connections for each AP in the network, $R_{i}$ is the achievable data rate of the AP in a certain connection.

			In \cite{Obeed_2018_Jour_power_allocation}, a new iterative joint power allocation and load balancing algorithm in a hybrid RF/VLC network for data rate maximization and system fairness improvement has been proposed. An iterative algorithm distributes users to APs and  the powers of the APs to their users.
			In \cite{Jin_2015_Jour_Delay_Guarantee_Femtocell}, optimal resource-allocation of mobile terminals in a heterogeneous wireless network under diverse QoS requirements has been considered. Decentralized algorithms were proposed for the resource-allocation problem. In \cite{Papanikolaou_2018_Conf_NOMA,Papanikolaou_2019_User_NOMA_Journal}, a NOMA hybrid RF/VLC with multiple VLC APs and one RF AP system has been investigated. In the proposed system both VLC and RF perform NOMA, and coalitional game theory is adopted for grouping the users.
			In \cite{Li_2015_Cooperative_Journal}, the average throughput of users for different FOV and LOS blocking probabilities is analyzed first, and then the  fairness from the perspective of systems and individual users is investigated to characterize the QoS experienced by users under different cell formation scenarios.

			\item  \textbf{Handover overhead:}
		
			Handover overhead is the control data sent over the network to coordinate the handover process between the control unit and users. This control signal is important to ensure that handover between RF and VLC APs is as smooth and fast as possible but it decreases the network efficiency and users' data rates.
            Therefore, it is critical to minimize the control data overhead. Handover overhead ($L_{HO}$) can be calculated as follows:
            \begin{equation}
            \label{eq:Handover_overhead}
            L_{HO} = \frac{R_{overhead}}{R_{data}},
            \end{equation}
            where $R_{overhead}$ is the average data rate in network for control signal and handover, and $R_{data}$ is the average data rate for the users' actual data.

            In \cite{Bao_2018_Conf_QoE_VHO}, on-off based vertical handover for a hybrid RF/VLC system that decreases delay and hence decreases the quality of experience penalty has been proposed. The proposed vertical handover is compared with I-VHO and D-VHO schemes. The simulation results reveal that the proposed scheme presents around 87.6\% drop in average handover delay cost and outperforms both I-VHO and D-VHO.  In \cite{Liang_2015_Conf_LTE}, a new vertical handover algorithm for a hybrid RF/VLC system is presented. Different network parameters such as average interruption duration and bit rate to assess the abilities of the proposed vertical handover algorithm to handle signal blockage/overload are considered.
            
            In \cite{Bao_2014_Protocol_Journal}, a hybrid network model of VLC and OFDMA is proposed for VLC hotspots. Furthermore, a novel protocol is proposed combined with access, horizontal, and vertical handover mechanisms for the mobile terminal (MT) to resolve user mobility among different hotspots and OFDMA systems.
				
			\item \textbf{Energy and power efficiency:}
			
			A lot of research in the wireless communication domain has studied techniques to reduce and minimize the networks' power consumption. However, it is important to relate the network's power consumption to average data rate and network outage and capacity. Therefore, calculating the power efficiency of the hybrid network is critical to compare it to standalone RF and VLC networks. Power efficiency ($\eta_{power}$) is calculated as the following\cite{Kashef_2016_Energy_Journal}:
            \begin{equation}
            \label{eq:Power_efficiency}
            \eta_{power} = \frac{P_{avg}}{R_{avg}},
            \end{equation}
            where $P_{avg}$ is the average power consumption of the network, and $R_{avg}$ is the average data rate of the users in the network.

            In \cite{Kashef_2016_Energy_Journal}, Energy efficiency was optimized in a heterogeneous RF/VLC network with multiple constraints on the minimum data rate for users and maximum power consumption per AP. It was shown how hybrid RF/VLC networks are more power-efficient than standalone RF networks. Moreover, the impact of multiple parameters such as LOS availability, the number of LEDs and the number of users in the network were taken into account while optimizing the power consumption.
            In \cite{Wu_2018_Jour_adap_resou_opti}, the network resource optimization problem for reducing communications power consumption while minimizing the queue lengths was investigated for hybrid RF/VLC networks.
            In \cite{Khreishah_2018_jour_Energy_Efficient}, an energy-efficient hybrid RF/VLC system for wireless access networks was investigated. The optimization problem is formulated by minimizing the power consumption subject to satisfying the users’ requests and maintaining an acceptable illumination level. In \cite{Hsiao_2018_Conf_Ener_Eff_Max}, a heterogeneous cellular network that combines RF and VLC to maximize the energy efficiency of the entire communication system under QoS requirements was considered. 

			\item \textbf{Area spectral efficiency (ASE):} 
			
			Although VLC networks offer a larger free spectrum over RF networks, VLC cells cover a smaller area than RF cells. Therefore, to be fair, when comparing the performance of hybrid networks with standalone RF and VLC networks, measuring area spectral efficiency (ASE) is needed. ASE ($\eta_{spectrum}$) measures the spectral efficiency per network (cell) area and is calculated as follows:  
             \begin{equation}
            \label{eq:Area_spectral_efficiency}
            \eta_{spectrum} = \frac{R_{avg}}{A_{cov}},
            \end{equation}
            where $R_{avg}$ is the average data rate of the users in the network and $A_{cov}$ is the area covered by the network with a certain minimum data rate.

			In \cite{Stefan_2014_Hybrid_Conf}, a scheme for cell association based
			on the minimum distance to the closest AP is proposed, where the ASE of a three-tier (macro-, femto- and optical attocells) 
			heterogeneous network is analyzed.
            It is shown that the average ASE of the hybrid RF/VLC the system can  be increased by at least two orders of magnitude
            over the stand-alone RF network.

			\item \textbf{Secrecy outage probability (SOP) and  achieved secrecy rate:}
			
			Secrecy is another important parameter that measures how the network design protects users from eavesdropping from any intruder in the network. Usually, VLC networks offer better security performance than RF networks because of the nature of the signal and the capability of the RF signal to penetrate walls, unlike visible light. Therefore, it is important to calculate the secrecy of the network to be able to compare the performance of the hybrid RF/VLC networks with standalone networks. 
			Considering a typical Wyner’s three-node model, there is a source (S) transmitting a secret message to a destination (D), while an eavesdropper (E) is trying intercept the information sent from S to D. If a silent (realistic) eavesdropping scenario is considered then S does not have any channel state information for the link between S and E. S will be sending with a constant rate of confidential message ($R_s$). The secrecy outage probability (SOP) is defined as the probability that the secrecy capacity ($C_s$) is less then $R_s$ \cite{Zhao_secrecy_SOP_2019_letter}, where $C_s = \max\{\log_2 (1 + \gamma_D) - \log_2(1 + \gamma_E),0\}$, and $\gamma_D$ and $\gamma_E$ are the instantaneous SNRs at D and E, respectively. Therefore, SOP ($P_{SOP}$) can be calculated as follows \cite{Bloch_secrecy_SOP_2008_Journal}:  
			\begin{equation}
            \label{eq:Secrecy_outage}
            P_{SOP} =  {P}\left[\log_2\left(1 + \gamma_D\right) - \log_2\left(1 + \gamma_E\right) > R_s\right].
            \end{equation}

			 In \cite{Khori_2019_Conf_Secrecy_Two_Stages,khori_2018_Conf_Phys_Lay_Secu}, the achievable secrecy capacity of DF-based hybrid RF/VLC was investigated. This paper considers two stages for achieving the required secrecy capacity. In the first stage, the beamforming vectors for  RF and VLC subsystems that maximize the achievable secrecy capacity are obtained. In the second stage, the power minimization algorithm that satisfies the required secrecy capacity is adopted.  
            In \cite{Khori_2019_Jour_Beamforming_Relaying}, the secrecy performance of relay-jammer selection/beamforming hybrid RF/VLC has been investigated assuming the absence of a direct link between source and destination. The considered system presents multiple DF relays and the relay node is selected by minimizing the outage probability. Jamming node is then selected from the available relaying nodes based on the received SNR at the eavesdropper location. Finally, the beamforming vectors for both RF and VLC subsystems are derived and used in the formulation of a power minimization task. 
            
            In \cite{Liao_2018_Jour_Physical_Layer_Security}, the secrecy performance of a hybrid RF/VLC system has been investigated where the relay node extracts the DC component and collects energy from the optical signal and then uses the collected energy for retransmitting data. Exact and asymptotic expressions are derived for secure outage probability and average secrecy capacity considering the effect of system parameters.  In \cite{Khori_2019_journal_Secrecy}, the secrecy capacity of DF-based hybrid RF/VLC has been compared with standalone RF and VLC systems and showed that the hybrid system exhibits better performance. The case of non-adaptive power allocation where both source and relay have the same amount of power and the case of cooperative power-saving where the total average power is shared between source and relay in a way that minimizes the total power while satisfying the required secrecy capacity were investigated.

		\end{itemize}

	\begin{table*}[!ht]
	\footnotesize
		\centering
\caption{Hybrid RF/VLC studies: Simulation and system implementation.}		
		\label{table:Implementation_Table}
		\linespread{1.15}\selectfont   

		\begin{tabular}{|>{\centering\arraybackslash}m{1.5cm}|>{\centering\arraybackslash}m{5cm}|>{\centering\arraybackslash}m{1.5cm}|>{\centering\arraybackslash}m{5cm}|}
			\hline
			\multicolumn{2}{|c|}{\textbf{Dual hop Hybrid RF/VLC}} & \multicolumn{2}{c|}{\textbf{Parallel Hybrid RF/VLC }} \\ \hline
			\textbf{References} & \textbf{Objective}  & \textbf{References}  & \textbf{Objective} \\ \hline
		\end{tabular}
		\begin{tabular}{|>{\centering\arraybackslash}m{1.5cm}|>{\arraybackslash}m{5.0cm}|>{\centering\arraybackslash}m{1.5cm}|>{\arraybackslash}m{5.0cm}|}
			\cite{Wentao_2017_Design_Journal} & Performance comparison of Hybrid RF/VLC system with standalone VLC and Wi-Fi networks in terms of coverage and network capacity. & \cite{Saud_2017_Conf_Software_Defined,Saud_2017_Conf_Software_Defined2}  & Implemented a hybrid RF/VLC communication system where VLC is considered as a primary link. The system monitors the VLC link and switches to RF if the VLC link fails. \\\hline

		 \cite{McBride_2014_Conf_flexible_testbed}& Based on the quality of first and second links, the system decides on the combination out of four possible combinations.  & \cite{Shao_2015_Design_Journal} & Proposed and implemented two heterogeneous RF/VLC systems. As a benchmark,  first a duplex system is investigated with VLC and RF links  for down- and  up-link, respectively. The bonding technique is applied to use the VLC and RF links in parallel and compare it with the benchmarking system in terms of robustness, throughput and average system delay. \\\hline

		\cite{Li_2018_Conf_BW_Aggreg}& Two hybrid dual hop RF/VLC systems were investigated:\begin{enumerate} \item Uplink and downlink data are transferred from the user through the RF channel while the downlink data is transferred via the VLC channel from  a relay to the user. \item Two duplex links between the user and the AP have been implemented.\end{enumerate}   & \cite{Yan_2016_combination_Conf}  & Implemented four different combinations: \begin{enumerate}
		 \item VLC and RF are used for both up- and down-links, and  RF is used if the VLC link drops.
\item VLC is used for both up- and down-links while RF is just for  uplink.  RF is used for uplink if the VLC link drops.
\item VLC is used for downlink and Wi-Fi is used for both up- and down-links. RF is used for downlink if the VLC link drops. 
\item VLC and RF are used, respectively,for down- and up-links.
		 \end{enumerate}\\\hline

		\end{tabular}
	\end{table*}


	\subsection{Hybrid Network Experimental Implementation}
\label{Sec:Hybrid_Network_Exp_implementation}

  Even though the literature is full of simulation-based VLC systems, there is a very limited number of researches for VLC system that consider the implementation of real-time VLC. However, for paving the way to commercial VLC products and for simulation-based system verification, the implementation under real-life conditions is an important step.
  The early proof of hardware implementation was demonstrated in \cite{Elgala_2007_VLC_OFDM,Elgala_2009_Indoor_Journal}, where the bandwidth of just 45 kHz was limited by the used digital signal processing (DSP) kit. The aim of the presented system was to investigate the performance of phase incoherent optical OFDM under different electrical SNRs. While in \cite{Elgala_2007_VLC_OFDM}, the considered system was unidirectional OFDM-based real-time experiments. In \cite{Elgala_2009_Indoor_Journal}, an array of nine white LEDs was used in an effort to enlarge the coverage area for OFDM experiments \cite{Elgala_2009_Indoor_Journal}. As a further improvement, a system achieving 100 Mbps has been implemented in \cite{Vuvcic_2010_DMT}, where the VLC- discrete multitone was considered. 
  The bottleneck for VLC implementation includes the LEDs driving circuits. The bias-T is the common LEDs driver that has been widely used as VLC's modulator due to its on shelves availability and can transmit OFDM signal with high PAPR. The function of bias-T is to add a high-frequency data signal on the top of DC signal. The main limitation of bias-T is its cost. In other words, installing indoor VLC systems with a huge implementation of bias-T is not practically possible. Therefore, seven DIP Op-Amps LED driving circuits has been recently proposed in \cite{Fuada_LED_driver_2018_Journal}.
  The other challenge is the widely used commercial white LED for indoor lighting fixtures. This type of lighting fixture generates light by exciting the yellow fluorescent powder by blue LED. While this lighting fixture has a long lifetime, it has a narrow bandwidth. Therefore, these have become the challenges of raising the VLC transmitting rate. Several methods have been used to improve the modulation bandwidth of the LED such as quantum dots (QDs) white LEDs, which were designed to improve the transmitting rate due to its high modulation bandwidths.
  One of the important aspects in VLC links is the received SNR which is affected by receiver's FOV and the LED's beam pattern. In an effort to improve the received SNR, the use of different lens combinations has been investigated in \cite{Aly_Experimental_VLC_2020_Conf} where Planoconvex lenses were used at the transmitter side to collimate the light pattern, and plano-convex and biconvex lenses were used at the receiver side to reduce the photodetector FOV and consequently enhance the receive SNR.

\begin{figure}[t!]
		\includegraphics[scale = 0.55]{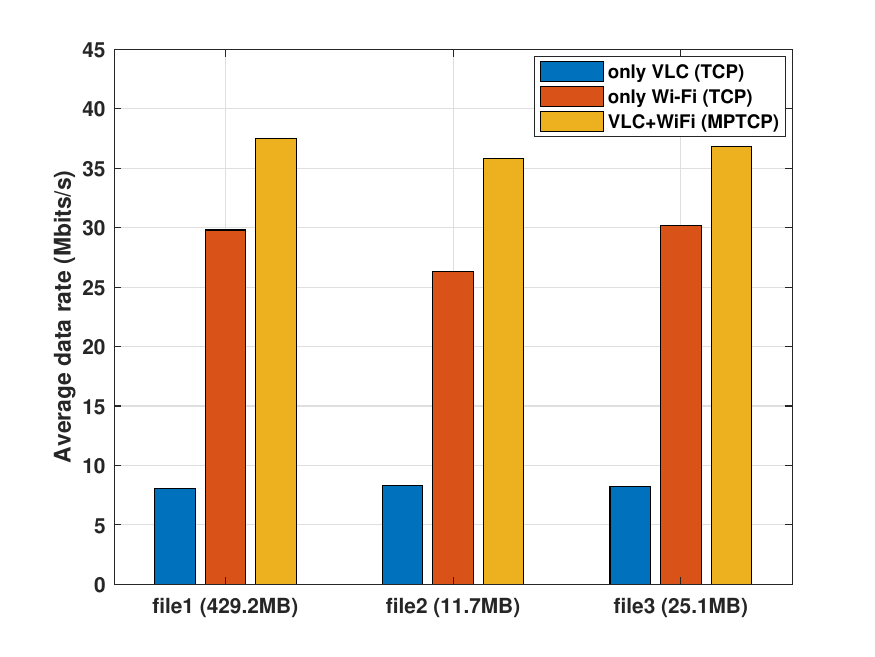}
			\centering
			\caption{Performance results for multi-link parallel transmission (Figure regenerated from \cite{Wentao_2017_Design_Journal}).}
			\label{fig:Result_Exp_hybrid1}
	\end{figure}

  Some studies have addressed experimental implementation of  hybrid systems such as  \cite{Yan_2016_combination_Conf,Shao_2015_Design_Journal,Wentao_2017_Design_Journal,McBride_2014_Conf_flexible_testbed,Saud_2017_Conf_Software_Defined,Saud_2017_Conf_Software_Defined2,Li_2018_Conf_BW_Aggreg}. In~\cite{Shao_2015_Design_Journal}, a proof of concept of coexistence of VLC and Wi-Fi networks was implemented. Two network implementations were proposed: the first uses the VLC channel as downlink and the Wi-Fi channel as an uplink, while the second aggregates Wi-Fi and VLC in parallel by using bonding technique in Linux OS. It turns out that the hybrid system improves the network throughput and decreases web page loading time compared to normal Wi-Fi networks. In~\cite{Wentao_2017_Design_Journal}, a hybrid network experiment platform was implemented. This reference shows that a  Wi-Fi/VLC hybrid network presents better performance than standalone VLC and Wi-Fi networks in terms of coverage and network capacity. In Fig.~\ref{fig:Result_Exp_hybrid1}, authors present the transmission rate of three different files using three different links only Wi-Fi, only VLC and parallel Wi-Fi/VLC. As shown in figure the results show that the hybrid approach improves the capacity gain of the network\footnote{The most important contribution of RF in a hybrid RF/VLC system is the stability. In VLC, weather including fog and rain can severely affect the link rate. Also, sun and its reflection can saturate the photodetectors\cite{bassam_Sun}. Further, there can be cases that objects block the line-of-sight, hence reduce the data rate in VLC drastically. There are studies which aim to overcome these challenges by proposing omnidirectional coverage in VLC \cite{Hossien_coverage}.}.
  
  In~\cite{McBride_2014_Conf_flexible_testbed}, possible benefits of hybrid RF/VLC such as reliable improvements of connectivity, throughput, coverage, and energy efficiency have been discussed.  In~\cite{Saud_2017_Conf_Software_Defined,Saud_2017_Conf_Software_Defined2}, a hybrid RF/VLC network that is capable of switching with high speed between VLC and RF has been implemented. The proposed system was validated by capturing video streams with low latency. In~\cite{Li_2018_Conf_BW_Aggreg}, a practical simultaneous two-link system is presented.  One of the links is a duplex RF link and the other one is an asymmetric duplex RF/VLC link. It turns out that the proposed system outperforms a standalone RF system as well as an asymmetric system. Studies of hybrid network simulation and system implementation are summarized in Table~\ref{table:Implementation_Table}. 
  While most of the studies used computer simulations, some other studies proposed using the ns3 network simulator as a reference, as discussed in \cite{Aldalbahi_2017_NS3_VLC}.

	\begin{table*}[ht]\AddFive
		\small
		\centering
		\caption{Application or Environment used in the different Hybrid Studies.}
		\label{table:Different_Environment_Application_Table}
		\renewcommand{\arraystretch}{1.15}
		\begin{tabular}{|>{\centering\arraybackslash}m{9cm}|>{\centering\arraybackslash}m{5cm}|}
			\hline
			\textbf{Application or  Environment need for Hybrid System}                                              & \textbf{References}
		\end{tabular}
		\begin{tabular}{|>{\arraybackslash}m{9cm}|>{\arraybackslash}m{5cm}|}
			\hline
			Energy efficient system  & $\!\!${\cite{Tran_2019_Ultra_Jorunal,Kong_2019_Energy_Jorunal,Rakia_2016_Optimal_DualHop_Journal,Rakia_2016_DualHop_VLC_RF_Conf,Kashef_2016_Energy_Journal,Zenaidi_2017_Achievable_Conf,Wu_2017_Dynamic_Conf,Obeed_2017_Joint_Conf,Pan_2017_3DHybrid_Conf,Kafafy_2017_Power_Conf,Pan_2017_Secure_Journal,Zhang_2018_Energy_Journal,Wu_2018_Jour_adap_resou_opti,Obeed_2018_Jour_power_allocation,Hsiao_2018_Conf_Ener_Eff_Max,Khreishah_2018_jour_Energy_Efficient,Zhou_2018_Conf_Coop_NOMA,Hammadi_NonOrthogonal_2021_Journal,Xiao_Cooperative_2021_Journal,Obeed_Power_2021_letter}} \\\hline

			RF sensitive indoor environment (hospitals)  & $\!\!${\cite{Hussain_2014_Hybrid_Conf,Vats_2017_Hybrid_Heath_Journal,Vats_2017_Modeling_Heath_Conf,Vats_2017_Conf_Outage_E_health_AF}} \\\hline

			Improving user data rate and capacity for indoor environment  & $\!\!${\cite{Chowdhury_2019_Integrated_Journal,Ma_Location_Conf,Papanikolaou_2019_User_NOMA_Journal,Basnayaka_2015_Hybrid_Conf,Wu_2016_Two_stage_Conf,Wu_2017_Access_Journal,Rahaim_2011_hybrid_Conf,Wu_2017_Access_Conf, Tabassum_2018_Coverage_Journal,Wang_2018_Learning_Journal,Ashimbayeva_2017_Conf_Hard_Soft_Switching,Zhang_2018_Conf_Coop_WSNs,Namdar_2018_jour_Outage_relay,Han_2018_Conf_Bipartite_Matching,Papanikolaou_2018_Conf_NOMA,Pratama_2018_Jour_BW_Aggreg,Amjad_Optimization_2020_Journal,Papanikolaou_Optimal_2020_Journal,Zeng_Realistic_2020_Journal,Hammadi_NonOrthogonal_2021_Journal,Liu_Optimization_2021_Conf,Aboagye_Joint_2021_Journal}}  \\\hline

			Improving avg. system delay for indoor environment  & $\!\!${\cite{Rakia_2016_DualHop_VLC_RF_Conf,Shao_2016_Delay_Journal,Hammouda_2018_Link_Journal,Jin_2015_Jour_Delay_Guarantee_Femtocell,Hammouda_2018_Conf_QoS_Constr,Bao_2018_Conf_QoE_VHO,KUCUK_selfadaptive_2021_Journal,Arshad_Stochastic_2021_Journal}} \\\hline

			High mobility environment  & $\!\!${\cite{Chowdhury_2019_Integrated_Journal,Li_2016_Mobility_Letter,Wang_2015_Dynamic_Journal,Wang_2015_Efficient_Journal,Wu_2017_Joint_Conf,Wang_2017_Optimization_Journal,Bao_2017_Visible_HetNet_Journal,Chowdhury_2014_Cooperative_Journal,Duvnjak_2015_Heterogeneous,Liang_2015_Conf_LTE,Zhang_2018_Jour_Spatially_Random,Du_2018_Jour_Knowledge_Transfer,Chen_Coordination_2021_Journal}}  \\\hline

			High secrecy physical-layer system  & $\!\!${\cite{Marzban_2017_Beamforming_Conf,Pan_2017_Secrecy_Conf,Pan_2017_Secure_Journal,Liao_2018_Jour_Physical_Layer_Security,khori_2018_Conf_Phys_Lay_Secu,Khori_2019_Jour_Beamforming_Relaying,Khori_2019_journal_Secrecy,Khori_2019_Conf_Secrecy_Two_Stages,Kumar_PLS_Analysis_2020_Journal}} \\\hline

			Improving uplink for VLC indoor systems  & $\!\!${\cite{Shao_2014_Indoor_Conf,Li_2015_Cooperative_Journal,Bao_2014_Protocol_Journal}} \\\hline

			IoT applications (high number of users and high data rate)  & $\!\!${\cite{Yang_2020_Learning_IoT_Journal,Wang_2015_Distributed_Conf,Mach_2017_Conf_D2D,Pan_2019_Jour_IoT,Becvar_2018_Conf_D2D,Wu_Data_Driven_2020_Journal,Peng_EndtoEnd_2021_Journal}}     \\\hline

			Vehicular application  & $\!\!${\cite{Bazzi_2016_Visible_Journal,Nauryzbayev_Outage_2020_Journal,Abouzohri_Performance_2020_Conf,Zadobrischi_System_Prototype_2020_Conf,Chen_Resource_2020_Letter,Chen_Coordination_2021_Journal}}  \\\hline

			Integrating PLC systems  & $\!\!${\cite{Kashef_2015_Achievable_Conf,Kashef_2016_Impact_PLC_Conf,Kashef_2017_Transmit_PLC_Journal}} \\\hline

			Demo testbeds (experimental implementation)  & $\!\!${\cite{Yan_2016_combination_Conf,Shao_2015_Design_Journal,Wentao_2017_Design_Journal,McBride_2014_Conf_flexible_testbed,Saud_2017_Conf_Software_Defined,Saud_2017_Conf_Software_Defined2,Li_2018_Conf_BW_Aggreg}} \\\hline
			
		\end{tabular}
	\end{table*}

	\subsection{Current Applications for Hybrid Systems}

    Hybrid RF/VLC can be classified based on the working environments. Hybrid RF/VLC is more feasible for the indoor environment due to the existence of both RF and VLC sub-systems. Therefore, most research works focus on indoor environments\cite{Hammouda_2018_Link_Journal}. Hybrid RF/VLC can also be used in outdoor environments. For example, in \cite{Ucar_2018_jour_outdoor}, a hybrid RF/VLC was proposed for a vehicular communication system.  A security protocol (SP) for vehicular communication to ensure platoon stability and maneuver under data packet injection, jamming, channel overhearing, and maneuver attacks was proposed in \cite{Ucar_2018_jour_outdoor}. 
    
    Hybrid systems were used by many studies for different environments and applications which are summarized in Table~\ref{table:Different_Environment_Application_Table}. 
    
    Light energy-efficient systems represent an important application. Many studies used proposed RF/VLC systems  for VL energy harvesting to improve the energy consumption of the hybrid system~\cite{Tran_2019_Ultra_Jorunal,Kong_2019_Energy_Jorunal,Rakia_2016_Optimal_DualHop_Journal,Rakia_2016_DualHop_VLC_RF_Conf,Kashef_2016_Energy_Journal,Zenaidi_2017_Achievable_Conf,Wu_2017_Dynamic_Conf,Obeed_2017_Joint_Conf,Pan_2017_3DHybrid_Conf,Kafafy_2017_Power_Conf,Pan_2017_Secure_Journal,Zhang_2018_Energy_Journal,Wu_2018_Jour_adap_resou_opti,Obeed_2018_Jour_power_allocation,Hsiao_2018_Conf_Ener_Eff_Max,Khreishah_2018_jour_Energy_Efficient,Zhou_2018_Conf_Coop_NOMA}.
    In~\cite{Kashef_2016_Energy_Journal}, the energy efficiency of a heterogeneous network of VLC and RF systems was investigated and numerical results corroborate the improvement in energy consumption of the proposed hybrid system. In~\cite{Zhou_2018_Conf_Coop_NOMA}, the authors proposed a novel cooperative non-orthogonal division multiple access methods that allows the hybrid RF/VLC network to simultaneously transmit wireless information and power.
    
    RF-sensitive indoor environments (e.g., hospitals) represent a critical application for VLC systems since the use of RF systems is limited due to the sensitivity of some RF-devices to RF-interference.  In~\cite{Hussain_2014_Hybrid_Conf,Vats_2017_Hybrid_Heath_Journal,Vats_2017_Modeling_Heath_Conf,Vats_2017_Conf_Outage_E_health_AF}, authors focused on using VLC network to decrease the usage of the RF network. In~\cite{Hussain_2014_Hybrid_Conf}, VLC was used to provide a downlink connectivity in RF prone environment.
    
    Improving user data rate and system capacity for the indoor environment is an important functionality of hybrid RF/VLC systems as shown in~\cite{Chowdhury_2019_Integrated_Journal,Ma_Location_Conf,Papanikolaou_2019_User_NOMA_Journal,Basnayaka_2015_Hybrid_Conf,Wu_2016_Two_stage_Conf,Wu_2017_Access_Journal,Rahaim_2011_hybrid_Conf,Wu_2017_Access_Conf, Tabassum_2018_Coverage_Journal,Wang_2018_Learning_Journal,Ashimbayeva_2017_Conf_Hard_Soft_Switching,Zhang_2018_Conf_Coop_WSNs,Namdar_2018_jour_Outage_relay,Han_2018_Conf_Bipartite_Matching,Papanikolaou_2018_Conf_NOMA,Pratama_2018_Jour_BW_Aggreg}. This is due  to the different characteristics of VLC and RF networks. Most of the studies optimize the resources of VLC and RF networks to maximize the users' data rate and average system capacity in an indoor environment. The same approach was used in~\cite{Rakia_2016_DualHop_VLC_RF_Conf,Shao_2016_Delay_Journal,Hammouda_2018_Link_Journal} but to minimize the average system delay. 
    
    Since high mobility indoor environments can be challenging for standalone VLC networks, studies  such as~\cite{Chowdhury_2019_Integrated_Journal,Li_2016_Mobility_Letter,Wang_2015_Dynamic_Journal,Wang_2015_Efficient_Journal,Wu_2017_Joint_Conf,Wang_2017_Optimization_Journal,Bao_2017_Visible_HetNet_Journal,Chowdhury_2014_Cooperative_Journal,Duvnjak_2015_Heterogeneous,Liang_2015_Conf_LTE,Zhang_2018_Jour_Spatially_Random,Du_2018_Jour_Knowledge_Transfer} proposed using optimizing hybrid RF/VLC network handover protocols to maximize the user connectivity and the overall network throughput. In~\cite{Li_2016_Mobility_Letter}, authors introduced a mobility aware load balancing scheme by leveraging the location sensitivity feature of the VLC network to decrease the overhead needed for handover and to increase the network throughput. 
    
    Taking advantage of the small coverage of VLC networks and the low penetration features (high physical layer security feature), studies such as \cite{Marzban_2017_Beamforming_Conf,Pan_2017_Secrecy_Conf,Pan_2017_Secure_Journal,Liao_2018_Jour_Physical_Layer_Security,khori_2018_Conf_Phys_Lay_Secu,Khori_2019_Jour_Beamforming_Relaying,Khori_2019_journal_Secrecy,Khori_2019_Conf_Secrecy_Two_Stages} used a hybrid network of RF/VLC systems to increase the PHY security of the network. In \cite{Marzban_2017_Beamforming_Conf}, the average consumed power by the hybrid RF/VLC network was minimized given a certain required secrecy rate via power allocation and beamforming algorithms.  Reference \cite{Khori_2019_journal_Secrecy}  compared the secrecy capacity of DF-based hybrid RF/VLC with standalone RF and VLC systems and showed that the hybrid system yields better performance. The case of non-adaptive power allocation where both source and relay present the same amount of power and the case of cooperative power-saving where the total average power is shared between source and relay in a way that minimizes the total power while satisfying the required secrecy capacity were both investigated. 
    
    Hybrid networks can be used to improve up-link for VLC indoor systems, a fact that was discussed in~\cite{Shao_2014_Indoor_Conf,Li_2015_Cooperative_Journal,Bao_2014_Protocol_Journal}. One of the early studies to present the idea of using both VLC and RF in the same network was~\cite{Shao_2014_Indoor_Conf}. An asymmetric RF/VLC combination was presented where VLC is used for downlink and RF is used for uplink to implement a full-duplex communication system.
    
    IoT and device-to-device (D2D) communication applications were addressed in~\cite{Yang_2020_Learning_IoT_Journal,Wang_2015_Distributed_Conf,Mach_2017_Conf_D2D,Pan_2019_Jour_IoT,Becvar_2018_Conf_D2D,Wu_Data_Driven_2020_Journal}.  In \cite{Wang_2015_Distributed_Conf}, authors introduced a new evolutionary game theory (EGT) algorithm that allows the users to adapt their needs to the hybrid RF and Li-Fi network.  
    
    Vehicular communications is a very promising application for hybrid systems because of its dynamic and changing features. In~\cite{Bazzi_2016_Visible_Journal}, the authors discussed the limitations of vehicular VL-only networks and the possibilities of utilizing them as a complementary technology with other RF wireless standards to improve the performance of conventional vehicular networks.
    
    Since power lines can be an efficient way to connect the VLC with the internet,  integrating power-line communication (PLC) systems with hybrid VLC and RF networks was discussed in~\cite{Kashef_2015_Achievable_Conf,Kashef_2016_Impact_PLC_Conf,Kashef_2017_Transmit_PLC_Journal}. In~\cite{Kashef_2016_Impact_PLC_Conf}, the impact of OFDM-based PLC backhauling in a multi-user hybrid system was discussed. Optimal power and subcarrier allocation algorithms were utilized to maximize the users' data rates. The hybrid system performance was studied numerically as a function of several parameters including PLC transmission power, RF and VLC spectrum, and the number of mobile terminals and APs.

    Finally, as seen in Table~\ref{table:Hybrid_Ref_years}, most of the new studies focus on implementing hybrid systems while old studies focus more on performance analyses and simulations.

	\begin{table*}
		\linespread{1.3}
		\footnotesize
		\centering

		\begin{center}
		\caption{Hybrid RF/VLC studies (chronologically ordered).}
			\label{table:Hybrid_Ref_years}
			\begin{tabular}{|>{\centering\arraybackslash}m{1.5cm}|>{\centering\arraybackslash}m{1.5cm}|>{\centering\arraybackslash}m{10.5cm}|}
				
				\hline
				\textbf{Year}  & \textbf{References} & \textbf{Comments (contributions)}
				\\  \hline
			\end{tabular}
			\begin{tabular}{|>{\centering\arraybackslash}m{1.5cm}|>{\centering\arraybackslash}m{1.5cm}|>{\arraybackslash}m{10.5cm}|}

                2021    & \cite{Xiao_Cooperative_2021_Journal} & 
			    A hybrid downlink system that proposes an optimization framework for VLC/RF information relaying and power transfer.
			    \\  \hline
                2021    & \cite{Chen_Coordination_2021_Journal} & 
			    Proposed an adaptive topology control scheme to achieve balance between connectivity and power consumption in VANET. 
			    \\  \hline
			    2021   & \cite{Hammadi_NonOrthogonal_2021_Journal} & 
			    Proposed an NOMA framework that is energy effiect for hybrid RF and VLC systems that take into account user mobility and imperfect channel state information.
			    \\  \hline
			    2021    & \cite{Obeed_Power_2021_letter} & 
			    Improves sum-rate and fairness of the networks by utilizing a cooperative NOMA scheme for power allocation and link selection in a hybrid RF/VLC system. 
			    \\  \hline
			    2021    & \cite{Peng_EndtoEnd_2021_Journal} & 
			    Proposed a dual hop system for optimized power and information transfer in IoT applications.
			    \\  \hline
	
			    2021    & \cite{Aboagye_Joint_2021_Journal} & 
			    Investigated the joint problem of AP assignment and power allocation in a hybrid RF/VLC system to maximize data rate and QoS under AP power constraint.
			    \\  \hline
			    2021    & \cite{Arshad_Stochastic_2021_Journal} & 
			    Conducted an user mobility analysis for RF/VLC hybrid networks by evaluating handover rates based on a random UE movement distribution to maximize network load and minimize handover rates.
s
			    \\  \hline
			    2020    & \cite{Wu_Data_Driven_2020_Journal} & 
			    Optimized the AP assignment in HetNet RF/VLC network for IoT applications ensuring QoS and minimum outage and handover overhead. 
			    \\  \hline
			    
			    2020    & \cite{Alenezi_Reinforcement_2020_Conf} & 
			    Improved fairness among all users and increase the overall network throughput by using reinforcement learning approach to improve network selection.
			    \\  \hline
			    
			    2020    & \cite{Nauryzbayev_Outage_Analysis_2020_Journal,Nauryzbayev_Outage_2020_Journal} & Studied outage Analysis of novel jamming-robust network for outdoor cognitive EV-enabled scenarios.

			    \\  \hline

			    2020    & \cite{Chen_Resource_2020_Letter} & Proposed a resource management algorithm for hybrid RF/VLC V2I wireless communication systems and demonstrated its energy efficiency enhancement. 
			    \\  \hline

			    2020    & \cite{Kumar_PLS_Analysis_2020_Journal} & Investigated physical layer security analysis for an indoor heterogeneous VLC/RF network along with introducing a novel secure-link selection mechanism based on known and unknown CSI. 
			    \\  \hline
			    
			    
			    2020    & \cite{Adnan_Load_Balancing_2020_Journal} & Proposed a dynamic cell formation method jointly with a load balancing strategy for hybrid VLC/RF networks based on blind interference alignment.
			    \\  \hline
			    
			    2020    & \cite{Amjad_Optimization_2020_Journal} &  Proposed an optimization scheme of joint problem of MAC frame slots and power allocation in hybrid VLC/RF network.
			    \\  \hline
			    
			    2020    & \cite{Papanikolaou_Optimal_2020_Journal} &  Proposed an optimal resource allocation for hybrid VLC/RF networks with common backhaul taking in consideration imperfect CSI.
			    \\  \hline
				
				2020    & \cite{Zeng_Realistic_2020_Journal} & Proposed a new resource allocation and load balancing scheme for WiFi and OFDMA-Based Li-Fi indoor hybrid  networks which improved the user average data rate.
			    \\  \hline
				
			    2020    & \cite{Rajo_jornal_2020} &  Implemented a a hybrid cellular architecture that allows inter-operability between different technologies (hybrid RF/VLC network) in the world of Internet of Things.
			    \\  \hline
			    
			    2020    & \cite{Yang_2020_Learning_IoT_Journal} & Presented an energy-efficient resource management for a heterogeneous RF/VLC in industrial IoT networks supporting different QoS requirements.
				\\  \hline
				2019    & \cite{Pan_2019_Jour_IoT} & Studied the outage performance of a hybrid VLC-RF indoor IoT system with light energy harvesting.
				\\  \hline

				2019    & \cite{Kong_2019_Energy_Jorunal} & Achieved lower outage probability using more energy efficient base station intensities for hybrid RF/VLC networks.
				\\  \hline
				
				2019    & \cite{Tran_2019_Ultra_Jorunal} & Proposed a collaborative resource allocation scheme for information and power transfer system in hybrid RF and VLC ultra cell networks. 
				\\  \hline

			    2018,2019    & \cite{Papanikolaou_2018_Conf_NOMA,Papanikolaou_2019_User_NOMA_Journal} & Proposed coalitional game approach for user grouping in hybrid RF/VLC networks With NOMA.
				\\  \hline

				2019    & \cite{Ma_Location_Conf} & Presented a location information-aided load balancing design for hybrid Li-Fi and Wi-Fi networks.
				\\  \hline

				2019    & \cite{Chowdhury_2019_Integrated_Journal} & Presented integrated RF/Optical wireless networks for improving QoS in indoor and transportation applications.
				\\  \hline


				\multicolumn{3}{r}{\footnotesize\textit{continued on the next page}}
			\end{tabular}
		\end{center}
	\end{table*}


	\begin{table*}
		\linespread{1.3}
		\footnotesize
		\centering
		\ContinuedFloat

		\begin{center}
		
				\caption{Hybrid RF/VLC studies (chronologically ordered).}

			\begin{tabular}{|>{\centering}m{1.5cm}|>{\centering}m{1.5cm}|>{\centering}m{10.5cm}|}
				\multicolumn{3}{l}{\footnotesize\textit{continued from previous page}}\\\hline  
				\textbf{Year}  & \textbf{References} & \textbf{Comments (Contributions)} 
			\end{tabular}

			\begin{tabular}{|>{\centering\arraybackslash}m{1.5cm}|>{\centering\arraybackslash}m{1.5cm}|>{\arraybackslash}m{10.5cm}|}\hline

				2019    & \cite{Khori_2019_journal_Secrecy} & Studied the secrecy capacity of decode and forward based hybrid RF/VLC relaying systems.
				\\  \hline
				
				2019   &	\cite{Khori_2019_Jour_Beamforming_Relaying}  &  	Investigated the secrecy performance of a design for a relaying hybrid RF/VLC system with minimized power. \\ \hline

				2018, 2017   &\cite{Becvar_2018_Conf_D2D, Mach_2017_Conf_D2D}&  Optimized selection between RF and VLC bands for device to device (D2D) communication in a multi user scenario to maximize spectral efficiency.\\ \hline

				2018   &\cite{Zhang_2018_Energy_Journal} & Investigated the energy-efficient subchannel and power allocation of heterogeneous VLC and RF based networks relying on the software-defined systems. \\ \hline

				2018   & \cite{Tabassum_2018_Coverage_Journal} &  Investigated coverage and rate analysis for RF/VLC downlink cellular networks for RF/VLC only, opportunistic hybrid RF/VLC, and centralized hybrid RF/VLC.  \\ \hline 
				

				2018   & \cite{Obeed_2018_Jour_power_allocation} &  {Proposed joint load balancing and power allocation for hybrid RF/VLC networks to improve system capacity and fairness.} \\ \hline 
				
				2018   & \cite{Zhang_2018_Jour_Spatially_Random} & {Investigated outage and symbol error probabilities of a hybrid RF/VLC under both DF and AF relaying schemes.} \\ \hline 
					
				2018   & \cite{Du_2018_Jour_Knowledge_Transfer} & {Presented a context-aware solution for indoor network selection using reinforcement learning based network algorithms.} \\ \hline

				2018   & 	\cite{khori_2018_Conf_Phys_Lay_Secu} &{Investigated beamforming design and  secrecy capacity of hybrid RF/VLC DF relaying systems in the presence of the eavesdropper nodes.} \\ \hline

				2018   &\cite{Hammouda_2018_Conf_QoS_Constr} & {{Studied effective capacity of a hybrid RF/VLC system subject to QoS constrains}} \\ \hline

				2018   & \cite{Namdar_2018_jour_Outage_relay} & Investigated outage and BER performances of AF relay-assisted hybrid RF/VLC system. \\ \hline
				
				2018   &\cite{Zhang_2018_Conf_Coop_WSNs} &  Investigated outage analysis of cooperative hybrid RF/VLC wireless sensor network (WSN) under DF and AF relay schemes. \\ \hline
				
				2018   &	\cite{Khreishah_2018_jour_Energy_Efficient} &  {Proposed a new paradigm in designing and reducing the power consumption of wireless indoor access networks satisfying user requests and needed illumination level.} \\ \hline

				2018   & \cite{Pratama_2018_Jour_BW_Aggreg} &{Proposed a bandwidth aggregation protocol for hybrid RF/VLC networks and implemented a real-life prototype of the hybrid VLC and RF communication system to test the protocol.} \\ \hline

				2018   & \cite{Liao_2018_Jour_Physical_Layer_Security} & {Studied the secure outage probability and average secrecy capacity of a dual-hop RF/VLC communication system.} \\ \hline 
				
				2018   & \cite{Zhou_2018_Conf_Coop_NOMA} &{Investigated outage performance of a cooperative NOMA based RF/VLC system.} \\ \hline

				2018   & \cite{Hsiao_2018_Conf_Ener_Eff_Max} & {Proposed coordinated beamforming design in a multiple input single output (MISO) downlink hybrid RF/VLC networks.} \\ \hline
				
				2018   & \cite{Wang_2018_Learning_Journal} &{Proposed learning aided network association for hybrid indoor Li-Fi and Wi-Fi Systems focusing on the LED AP selection strategies.} \\ \hline
				
				2017   & \cite{Wang_2017_Optimization_Journal} &{Investigated optimization of load balancing in hybrid RF/Li-Fi Networks.}  \\ \hline


				{2017, 2016}   &\cite{Wu_2016_Two_stage_Conf,Wu_2017_Access_Journal} & {Proposed an AP selection method for VLC and RF networks to improve the throughput of the network.} \\ \hline

				2017   & 	\cite{Kashef_2017_Transmit_PLC_Journal} &  {Analyzed the optimization problem of minimizing the total transmission power consumption for a hybrid power-line communication (PLC)/RF/VLC system.} \\ \hline

				2017   &\cite{Pan_2017_Secrecy_Conf,Pan_2017_Secure_Journal} & {Conducted secrecy outage analysis of hybrid RF/VLC system with light energy harvesting.} \\ \hline

				2017   &\cite{Wu_2017_HRO_OFDM_Journal} &  {Proposed a new OFDM design for hybrid RF/VLC baseband system consider both power and bandwidth allocation.} \\ \hline  
				
				2017   &\cite{Vats_2017_Hybrid_Heath_Journal,Vats_2017_Modeling_Heath_Conf} & {Studied modeling and outage analysis of hybrid RF/VLC system for real time health care applications outage probability.} \\ \hline 
			
				2017   & \cite{Wentao_2017_Design_Journal} &{Developed an indoor hybrid VLC-Wi-Fi network  network experiment platform to test coverage and network capacity.}  \\ \hline 
				
				
				2017   & \cite{Saud_2017_Conf_Software_Defined,Saud_2017_Conf_Software_Defined2} & {Implemented a software defined hybrid RF/VL network that is capable of switching with high speed between VLC and RF.} \\ \hline

			\multicolumn{3}{r}{\footnotesize\textit{continued on the next page}}
				
			\end{tabular}
		\end{center}
	\end{table*}

	\begin{table*}\AddFive
		\linespread{1.3}
		\ContinuedFloat
		\footnotesize

		\begin{center}
		
		\caption{Hybrid RF/VLC studies (chronologically ordered).}

			\begin{tabular}{|>{\centering}m{1.5cm}|>{\centering}m{1.5cm}|>{\centering}m{10.5cm}|}
				\multicolumn{3}{l}{\footnotesize\textit{continued from previous page}}\\\hline  
				\textbf{Year}  & \textbf{References} & \textbf{Comments (Contributions)}
			\end{tabular}
			\begin{tabular}{|>{\centering}m{1.50cm}|>{\centering}m{1.50cm}|>{}m{10.5cm}|} \hline

				2016   & \cite{Kashef_2016_Energy_Journal} & {Investigated and optimized power and bandwidth allocation for energy efficiency maximization of a heterogeneous RF/VLC network.} \\ \hline 
			
				2016 &\cite{Rakia_2016_Optimal_DualHop_Journal,Rakia_2016_DualHop_VLC_RF_Conf}  &  {Optimized energy harvesting and data rate under delay constraint in daulhop hybrid RF/VLC system.} \\ \hline
				
				2016   & \cite{Bazzi_2016_Visible_Journal} & {Introduced the use of VLC as a complementary technology for the DSRC for connected vehicles (Future of heterogeneous vehicular network).} \\ \hline  
				
				2016   & \cite{Li_2016_Mobility_Letter} & {Proposed a mobility aware load balancing in LTE/VLC networks to improve system throughput and handover overhead } \\ \hline 
				
               2016   & \cite{Shao_2016_Delay_Journal}  & {Investigated delay analysis of RF/VLC coexistence wireless networks for aggregated and non-aggregated scenarios.} \\ \hline

		
				2015   &\cite{Li_2015_Cooperative_Journal} & {Investigated cooperative load balancing in hybrid VLC and Wi-Fi to improve high area spectral efficiency and optimize system throughput.}\\ \hline 
				
				2015   &\cite{Wang_2015_Dynamic_Journal} & {Investigated dynamic load balancing with handover in hybrid Li-Fi and Wi-Fi networks  taking into  account user mobility and handover signaling overhead.} \\ \hline 
				
				2015   & \cite{Jin_2015_Jour_Delay_Guarantee_Femtocell} & {Studied resource allocation under delay-guarantee constraints for heterogeneous visible-Light and RF femtocell.} \\ \hline

				2015   &  \cite{Shao_2015_Design_Journal} & {Conducted design and analysis of a VLC enhanced Wi-Fi system with and without aggregation.} \\ \hline 
			    2015   & \cite{Rahaim_2015_Toward_Journal} & {Investigated practical integration of VLC in 5G networks where motivation of using VLC and system component needed in deployment.}\\ \hline 

			2015   &  \cite{Wang_2015_Efficient_Journal}  & {Proposed an efficient vertical handover scheme for RF/VLC systems using a Markov decision process to obtain a trade-off between switching cost and delay requirement.} \\ \hline


				2014   &  	\cite{Bao_2014_Protocol_Journal} & {Proposed a hybrid network of VLC and OFDMA system in which VLC is used only for downlink to improve the network capacity.} \\ \hline

				2014   &  \cite{Chowdhury_2014_Cooperative_Journal} & {Studied performance of cooperative data download indoor hybrid wireless local area networks (WLAN)-VLC and compared it with standalone WLAN-only and VLC-only networks.} \\ \hline 
				
				2014   &  \cite{Hussain_2014_Hybrid_Conf} & {Proposed the use of hybrid RF/VLC systems in RF sensitive indoor environments.} \\ \hline 
				
				2014   & \cite{McBride_2014_Conf_flexible_testbed} & {Implemented a Software defined hybrid RF/VLC system using off-the-shelf LEDs and photodetectors.} \\ \hline

				2011   &\cite{Rahaim_2011_hybrid_Conf}  & { Proposed an indoor hybrid system that integrates WiFi and VLC luminaries to improve throughput and delay for broadcasting.}\\ \hline  
				
			\end{tabular}\AddFive
		\end{center}
	\end{table*}

	\section{Research Directions and Future Work }
	\label{sec:Research_Directions}

This section discusses the open research directions and future work for the hybrid RF/VLC studies.

\subsection{Current Research Directions}
\begin{itemize}

			\item	\textbf{Energy harvesting with dual-hop RF/VLC system under delay and data rate constraints:}
			
Multiple studies have provided simulation analysis of a dual hop system consisting of VLC and RF links. 
While the idea of harvesting energy as well as transferring data between the user and APs is a new idea and it appears to be a power-efficient solution. The drawbacks of energy harvesting where delay is introduced have to be minimized.  Most of the studies have simulated the system and optimized it in terms of a single constraint like overall system delay, maximum user data rate, and system secrecy. Further analysis should consider multiple constraints and conduct more analysis for such systems. Moreover, a good handover scheme needs to be adopted to ensure system reliability. Finally, testing such system in a real hardware prototype would also be a possible future research direction.

             \item	\textbf{Resource allocations and AP selection in RF/VLC networks:}

Multiple studies have discussed the AP selection in RF/VLC networks. Depending on whether the hybrid network design has a central control unit or not, the algorithm will vary. The main goal is to find an energy-efficient algorithm that will ensure the fairness of the network over the VLC and RF networks and maximize the throughput of the users in the network. Many studies can be conducted in the direction of utilizing machine learning to solve this non-convex optimization problem. Moreover, future work would be focusing on using the most efficient and low-cost optimization algorithm to ensure the best network throughput, and minimizing the control overhead, delay, and power transmitted.  In addition, resource allocations such as bandwidth allocations and the number of APs would be critical network design problems that would depend on the requirement of the network, the availability of the resources, and mobility of the users. The metrics used to compare different algorithms are network fairness, user satisfaction, and outage probability.

            \item	\textbf{Load balancing and handover between VLC and RF networks:}

It is evident from the current research that hybrid RF/VLC networks provide better performance in terms of bandwidth and user mobility. However, it is important to maximize the benefits of both networks by applying an efficient load balancing scheme and a fast handover algorithm to ensure network reliability in all times and locations.  Researches are focusing on minimizing the handover latency and increasing the network overall throughput by using different techniques of load balancing schemes as mobility-aware and location-aware handover algorithms. The future direction for this challenging issue in hybrid networks is having a dynamic load balancing scheme that updates the network resources whenever needed. Moreover, utilizing distributed users' shared information is another solution to decrease control overhead and power transmitted by the users. Finally, different solutions could be compared with respect to user maximum data rate, control overhead, network fairness, and outage probability.

           \item 	\textbf{Physical layer security of RF/VLC networks:}

Secrecy is a growing network aspect and issue that is considered by multiple applications and users. Multiple studies have  focused on increasing the security of the network and measuring the probability of the network's sustainability to eavesdropping. Since VLC networks are superior to RF networks in this particular aspect, the effect of using hybrid RF/VLC networks is studied and simulated by multiple researchers. Because the location of the eavesdropper is an important parameter that affects the secrecy of the network, more analysis would be needed to ensure that the network power consumption is minimized while ensuring a certain level of secrecy for the networks in different scenarios and network topologies.

           \item  	\textbf{Design and implementation of practical hybrid RF/VLC network (testbeds):}
           
           The most important step of proposing a new algorithm is to test it on real hardware. Therefore, designing and implementing testbeds for hybrid RF/VLC networks is critical for the development of any valid comparison between proposed handover and resource allocation algorithms. Moreover, these practical implementations will vary in cost and number of users that it can support. In the future, further optimization methods for the currently proposed testbeds would be needed to make it easier to implement on available hardware and give more flexibility to the researchers in terms of network design and the number of users supported. The development of lower-cost testbeds will allow more studies to be done on hybrid networks optimization algorithms from load balancing to AP selection. In addition, the availability of easy-to-setup hardware will make it more accessible for researchers to develop new applications and use cases for hybrid RF/VLC networks.

           \item  	\textbf{New modulation schemes and access methods (OFDM, NOMA) for Hybrid RF/VLC networks:}

          Numerous studies have been conducted on testing new modulation techniques and access methods. Devices that support hybrid RF/VLC would require its internal hardware RF front-end and baseband blocks (to be updated and optimized). 
          For example, researchers in\cite{Wu_2017_HRO_OFDM_Journal} proposed a new hybrid of radio optical OFDM (HRO-OFDM) scheme which combines both RF and VLC link in the physical layer. This new OFDM scheme design improves optimization and allocation of both power and bandwidth for both technologies which allows the network to achieve better performance.  Moreover, multiple studies proposed using NOMA which appeared to be less sensitive to LOS availability compared to OFDMA\cite{Papanikolaou_2019_User_NOMA_Journal,Hammadi_2019_Robust_NOMA_Conf,Obeed_Journal_NOMA_User_Pairing_2020}. In \cite{Hammadi_2019_Robust_NOMA_Conf}, a novel energy efficient NOMA hybrid RF/VLC system has been proposed. Particularly, authors have considered using downlink hybrid RF/VLC and deriving a closed-form expression for energy efficiency which is subsequently used for extensive energy efficiency analysis of their proposed scheme. Authors have further demonstrated that NOMA is less sensitive to LOS compared to its OFDMA-based counterpart. In \cite{Obeed_Journal_NOMA_User_Pairing_2020}, multiple access techniques were performed for maximizing the overall achievable data rate where they jointly decide on the power allocation coefficients, pairing index, and link selection. However, the non-cooperative nature of NOMA makes users competing on resources. Therefore, studies proposed using a coalitional game approach instead of standard opportunistic scheme\cite{Papanikolaou_2019_User_NOMA_Journal}. In the future, it is expected that researchers would continue improving the modulation schemes and access methods utilized by hybrid RF/VLC networks to further improve the performance and optimize the utilized resources.

           \item 	\textbf{Environment adaptive RF/VLC networks (interference, data rate, coverage, number of users requirements) and application dependent networks:}
           
           As seen throughout the paper, hybrid RF/VLC networks were proposed to be utilized in different applications to make use of their advantages and flexibilities. However, depending on the application and its environment, either indoor or outdoor, the network parameters need to be adapted to fully utilize the benefits of a hybrid RF/VLC network. Moreover, it is expected that more applications would emerge for hybrid networks as more researchers get easy to use testbeds and realize the benefits and advantages of hybrid networks. This means new algorithms and network topologies would need to be introduced and studied to adapt networks to new environments and applications.

 \item  	\textbf{Backhauling of RF/VLC networks:}
           Backhauling is the connection of the APs of RF and VLC network to core network. Usually the AP side is the bottleneck of the network. However, with new modulation techniques and coding scheme the bottleneck of the network shifted to the connection between the AP and the core network as number of users and data rate demand increase. Some studies have discussed this issue and proposed solutions and analysis for the hybrid networks \cite{Papanikolaou_Optimal_2020_Journal,Kashef_2016_Impact_PLC_Conf,Kashef_2015_Achievable_Conf}. Another direction is the use of the dual hop network architecture as a solution where the VLC provides the backhaul for the RF APs instead of fiber cable which provides more flexibility to the network installation. More studies are needed to compare between the two solutions and provide more analysis on the costs and benefits.

	\end{itemize}
	
	\subsection{Expected Future Research Directions}

After discussing the research work carried out in the area of  hybrid RF/VLC networks, we present expected future directions in this section. From the trends we have seen in the VLC research, it is clear that more applications that are based on visible light technologies are going to be introduced in the future. More coexistence between RF networks and VLC based networks will be required to improve the user experience and reduce the networks delay and power consumption. Therefore, the following research directions could be investigated:

\begin{itemize}

        \item Introducing new novel and efficient optimization techniques to improve the performance of hybrid RF/VLC networks.
        
        \item Improving implementations for the proposed network configurations using software defined radio systems (SDRs).
        
        \item Improving the energy efficiency of the network.
        
        \item Performing spectrum sensing and utilizing the best available RF or visible light spectrum by applying machine learning algorithms (e.g., cognitive radio).
        
        \item Finding new applications for hybrid systems (emerging applications for the hybrid RF/VLC systems).
        
         As discussed throughout the paper, hybrid RF/VLC technology is very important as it provides both advantages of RF and VLC standalone systems. Therefore, it will add diversity, increase coverage, decrease power consumption and interference. Potential applications are listed as follows:
       
        \begin{enumerate}
 
\item   As a replacement for normal Wi-Fi systems, especially for indoor scenarios.
\item  In RF interference-free applications to decrease the RF power used, e.g., airports and hospitals.
\item In vehicular communication technologies as V2X to make use of the built-in hardware (hybrid DSRC/VLC).
\item Power efficient IoT applications and Machine-to-Machine (M2M) communication.
\item In sensing applications as hybrid RF/VLS (VLS: Visible Light Sensing),  e.g., occupancy estimation and positioning.

        \end{enumerate}
	\end{itemize}

	\section{Conclusions}
	\label{sec:conclusion}
	
	The idea of having a dynamic and reconfigurable network that can support both RF and VLC is being adopted by a lot of studies and researchers in recent years. This is due to the  multiple advantages that these systems possess:
    \begin{itemize}
        \item Having a stable channel under different circumstances.
        
        \item Improving user mobility and security.
        
        \item Making the network less RF interfering.
        
        \item Improving network power consumption, delay, and capacity.
        
        \item Improving network reliability under a high number of users and data rates.
        
        \item Integrating it easily with multiple technologies and in different applications.
    \end{itemize}
    
    These types of networks present features such as smart cognitive networks that are capable of adapting to the surrounding environment and exploiting optimally the available resources.
    These networks offer flexibility and high performance, which greatly suits the new requirements of future networks such as 5G and beyond.
    
    This paper summarized all the important aspects of these hybrid networks. First, we have shown how different network topologies can be used in these hybrid systems depending on the application considered.  These topologies are currently being tested in the recent hardware implementations of these hybrid networks. Second, we discussed the different optimization techniques to maximize the performance of these networks. Third, we described different performance analysis tools for hybrid networks. These tools ranged from network metrics to user specific metrics. Moreover, we presented the applications and environments where these hybrid networks find applicability and outperform the normal standalone networks. These applications are going to expand more as more studies dedicated to these hybrid networks.
    
    Finally, as a conclusion of the survey work, it is important to note that hybrid RF/VLC networks come in different network topologies, use variable handover and load balancing schemes, and utilize a wide range of modulations and access methods. The most important thing is to understand the nature of the intended application of each user-case for the network to adapt its specifications to the needs of the users and available resources. Future analyses and simulations for these hybrid networks are still needed for hybrid networks in terms of access methods and load balancing to take advantage of the idea of using two different technologies (operating frequencies) competing in the same network. Eventually, these networks would evolve into smarter cognitive networks that would maximize the performance of their users while maintaining efficient resource utilization.

	\section*{NOMENCLATURE}
	\begin{description}[align=left,leftmargin=2cm,font=\normalfont,style=nextline,itemsep=0pt]
		\item[ADC] Analog to Digital Converter
		\item[AF]  Amplify and Forward 
		\item[APs] Access Points 
		\item[AR] Augmented Reality
		\item[ASTM]  American Society for Testing and Materials 
		\item[BER]  Bit Error Rate 
		\item[bps]  Bits Per Seconds
		\item[BS] Base Station
		\item[BS] Base Station
		\item[CDMA] Code Division Multiple Access
		\item[CSK] Color Shift Keying 
		\item[CSMA] Carrier Sense Multiple Access
		\item[D2D] Device to Device
		\item[DAC] Digital to Analog Converter
		\item[DC] Direct Current 
		\item[DF] Decode and Forward 
		\item[DSRC] Dedicated Short Range Communications 
		\item[DSSS] Direct Sequence Spread Spectrum 
		\item[D-VHO] Dwell Vertical Handover  
		\item[EGT] Evolutionary Game Theory 
		\item[EM] Electromagnetic  
		\item[eU-OFDM] enhanced Unipolar Orthogonal Frequency Division Multiple Access
		\item[FHSS] Frequency-Hopping Spread Spectrum 
		\item[FOV] Field Of View  
		\item[FPS] frame per second
		\item[FSO] Free Space Optical
		\item[HetNet] Heterogeneous Network  
		\item[I2V] Infrastructure-to-Vehicle 
		\item[IM/DD] Intensity Modulation/Direct Detection  
		\item[IIoT] Industrial Networks Internet of Thing
		\item[IoT] Internet of Thing  
		\item[IR] Infrared Radiation 
		\item[ITS] Intelligent Transportation Systems 
		\item[I-VHO] Immediately executes Vertical Handover 
		\item[kbps] Kilobits Per Seconds 
		\item[LB] Load Balancing
		\item[LED] Light Emitting Diode  
		\item[Li-Fi] Light Fidelity 
		\item[LOS] Line of Sight  
		\item[LTE] Long Term Evolution  
		\item[M2M] Machine to Machine 
		\item[MAC] Medium Access Control 
		\item[MBE] Mean Bandwidth Efficiency
		\item[MBSs] Macro Cell BSs
		\item[MDP] Markov decision process
		\item[MGF] Moment Generating Function 
		\item[MIMO] Multiple Input Multiple Output
		\item[MISO] Multiple Input Single Output 
		\item[mmWave] millimeter Waves  
		\item[MOO] Multi-Objective Optimization 
		\item[MR] Mobile Robots
		\item[NOMA] Non-Orthogonal Multiple Access 
		\item[NLOS] Non-LOS 
		\item[NP] Nondeterministic Polynomial
		\item[OCC] Optical Camera Communication
		\item[OFDM] Orthogonal Frequency Division Modulation 
		\item[OFDMA] Orthogonal Frequency Division Multiple Access 
		\item[OWC] Optical Wireless Communications
		\item[PAPR] Peak-to-Average-Power Ratio 
		\item[PD] Photo-Detector 
		\item[PDF] Probability Density Function 
		\item[PDS-ERT] Post-Decision State-Based Experience Replay and Transfer
		\item[PHY] Physical Layer 
		\item[PLC] Power-Line Communication 
		\item[QoS] Quality of Service 
		\item[RF] Radio Frequency  
		\item[SDR] Software Defined Radio 
		\item[SNR] Signal-to-Noise Ratio 
		\item[SOO] Single-Objective Optimization 
		\item[SP] Security Protocol 
		\item[SBSs] Small Cell BSs
		\item[UE] User Equipment 
		\item[V2I] Vehicle-to-Infrastructure 
		\item[V2V] Vehicle-to-Vehicle 
		\item[VL] Visible Light  
		\item[VLC] Visible Light Communication  
	    \item[VLCBSs] VLC BSs
	    \item[VT] Vectored Transmission
	    \item[V2X] Vehicle-to-Everything
		\item[WAVE] Wireless Access in Vehicular Environments 
		\item[Wi-Fi] Wireless Fidelity 
		\item[WLAN] Wireless Local Area Networks 
		\item[WSN] Wireless Sensing Network 
\end{description}	
	
	\section*{Symbols}
	\AddFive\AddFive

	\begin{description}[align=left,leftmargin=2cm,font=\normalfont,style=nextline,itemsep=0pt]
		\item[$A_R$] Optical detector size  
		\item[$d$] Distance between the receiver and the transmitter  
		\item[$FOV$] Field of view for VLC receiver  
		\item[$\varsigma$] 	Channel path loss exponent  
		\item[$g(\theta)$] Concentrator gain  
		\item[$H_v$]  Optical channel gain  
		\item[$K$] All the gains and the transmitted power 
		\item[$m$] Order of the Lambertian model  
		\item[$\phi$] Irradiance angle 
		\item[$\phi_{\text{max}}$] Semi-angle at half-power of the LED 
		\item[$P_t$] Maximum allowable total power  
		\item[$P_{t-RF}$] Power transmitted for RF system  
		\item[$P_{t-VLC}$] Power transmitted for VLC system 
		\item[$\theta$] Incidence angle 
		\item[$T_s(\theta)$]  Gain of the optical filter  
		\item[$ R_{i} $]  Average data rate of user $i$ 
		\item[${E}\rm{ \left[ \cdot \right]}$ ]      Expected value operator 
		\item[$  \gamma_i(t) $]     SINR over time for user $i$
		\item[$ B $] User assigned bandwidth 
		\item[$ T^{i}_{tr}$] Multi-hop retransmission time delay per hop $i$
		\item[$ T_{err}$]  Error retransmission time delay
		\item[$ T_{HO} $]  Handover time delay  
		\item[$ T_{avg} $]  Average transmission delay 
		\item[$ {P}_{HO} $] Probability a handover process occurs 
		\item[$ {P}_{err} $] Probability a transmission error occurs 
		\item[$ N $] Maximum number of hops in the system 
		\item[$  BER $]   Bit error rate 
		\item[$ P_c $]  Signal coverage probability
		\item[$ \gamma_{o} $] Users' instantaneous SINR 
		\item[$ \gamma_{th} $]  Target SINR threshold  
		\item[$ F $]  Jain's fairness index 
		\item[$ N_{AP} $]  Number of connections for all APs in network
		\item[$ R_{i} $]  Achievable data rate of the AP per connection 
		\item[$ L_{HO} $]  Handover overhead 
		\item[$ R_{overhead} $]  Average data rate of control signal and handover 
		\item[$ R_{data} $] Average data rate for the users' actual data  
		\item[$ \eta_{power} $] Network power efficiency 
		\item[$ P_{avg} $]  Average power consumption of the network 
		\item[$R_{avg}$]  Average data rate of all users  
		\item[$ \eta_{spectrum} $] Area spectral efficiency  
		\item[$ A_{cov} $]  Area covered by the network with a minimum data rate 
		\item[$ R_s $]  Source constant data rate of confidential message 
		\item[$ P_{SOP} $]  Secrecy outage probability 
		\item[$ C_s $] Secrecy capacity  
		\item[$ \gamma_D $]  Instantaneous SINR at the destination 
		\item[$ \gamma_E $]  Instantaneous SINR at the eavesdropper 
\end{description}

\phantomsection
	\section*{Acknowledgment}
The authors would like to thank Dr. Qian Gao for his great support in the preparation of this study and his helpful suggestions and perspectives, and Md Zobaer Islam for his valuable comments and suggestions to improve this paper. 	
This publication was made possible by the NPRP 13S-0130-200200 from the Qatar National Research Fund (a member of The Qatar Foundation). The statements made herein are solely the responsibility of the authors.

	\bibliographystyle{IEEEtran}
	\bibliography{Bibliography.bib}

\end{document}